\pdfoutput=1
\newif\ifnotes

\PassOptionsToPackage{dvipsnames,table}{xcolor}
\PassOptionsToPackage{hyphens}{url}
\documentclass[conference]{IEEEtran}
\IEEEoverridecommandlockouts
\usepackage{xcolor}
\usepackage{amssymb}
\usepackage{cite}
\usepackage{url}
\usepackage[hidelinks]{hyperref}
\newcommand{\Description}[1]{}

\usepackage{tikz}
\usetikzlibrary{patterns, arrows.meta, positioning, decorations.pathreplacing, fit, shapes.geometric}
\usepackage{float}
\usepackage{algpseudocode}
\usepackage[normalem]{ulem}
\usepackage{setspace}
\usepackage{xcolor}
\usepackage{soul}
\ifnotes
  \usepackage[nomargin,inline,draft]{fixme}
\else
  \usepackage[nomargin,inline,final]{fixme}
\fi

\newif\ifrevisions
 \revisionsfalse
\ifrevisions
  \newcommand{\rev}[1]{\textcolor{blue}{#1}}
\else
  \newcommand{\rev}[1]{#1}
\fi
\definecolor{addcol}{RGB}{0,128,60}
\ifrevisions
  \newcommand{\add}[1]{\textcolor{addcol}{#1}}
\else
  \newcommand{\add}[1]{#1}
\fi
\definecolor{leancol}{RGB}{160,0,150}
\newif\ifleanrev
\leanrevfalse
\ifleanrev
  \newcommand{\lean}[1]{\textcolor{leancol}{#1}}
\else
  \newcommand{\lean}[1]{#1}
\fi

\usepackage[utf8]{inputenc}
\usepackage[T1]{fontenc}
\usepackage{microtype}
\usepackage{graphicx}
\usepackage{amsmath}
\usepackage{xspace}
\usepackage{listings}
\usepackage{multirow} \usepackage[most]{tcolorbox}
\usepackage{booktabs}
\usepackage{tabularx}
\usepackage{pifont}
\usepackage{amsthm}

\usepackage{mathtools}

\usepackage{relsize}

\usepackage{caption}
\usepackage{subcaption}
\usepackage{mdframed}

\newcommand{\pnameTitle}{\texttt{Pearl}}
\newcommand{\pname}{\pnameTitle}

\newcommand{\nameol}{\textnormal{\texttt{Pearl-OL}}}
\newcommand{\namebof}{\textnormal{\texttt{Pearl-BOF}}}

\definecolor{highlight1}{HTML}{F58F33} \definecolor{highlight2}{HTML}{2F7BA3} 

\colorlet{highlight1_transparent}{highlight1!50} \colorlet{highlight2_transparent}{highlight2!50}

\newtheorem{definition}{Definition}

\newcommand{\commentout}[1]{}

\definecolor{white}{RGB}{255,255,255}
\definecolor{vrpink}{RGB}{255,0,127}
\definecolor{vrblue}{RGB}{30,144,255}
\definecolor{vrolive}{RGB}{85,107,47}
\definecolor{vrroyalblue}{RGB}{65,105,225}
\definecolor{brgreen}{RGB}{100,200,70}
\definecolor{ivsalmon}{RGB}{255,160,122}
\definecolor{vrlpink}{RGB}{255,192,203}
\definecolor{mvcol}{RGB}{5,150,25}

\usepackage[obeyFinal]{todonotes}

\newcommand{\locprec}{\prec_{\mathrm{loc}}}

\fxusetheme{color}
\fxuseenvlayout{color}

\graphicspath{{./figures/}}
\DeclareGraphicsExtensions{.pdf,.jpeg,.png,.jpg}

\usepackage{comment}

\usepackage{algorithm}
\newcommand{\algfont}{\footnotesize}

\algblockdefx{As}{EndAs}[1]{\textbf{as a }#1}{\textbf{}}
\algblockdefx{All}{EndAll}[1]{\textbf{all replicas}#1}{\textbf{}}
\algblockdefx{Timeout}{EndTimeout}[1]{\textbf{upon timeout}#1}{\textbf{}}
\algblockdefx{Func}{EndFunc}[1]{\textbf{function }#1}{\textbf{}}
\algblockdefx{WFunc}{EndWFunc}[3]{\textbf{function }#1\textbf{ where }#2\textbf{ is }#3}{\textbf{}}
\algblockdefx{WFuncx}{EndWFuncx}[3]{\begin{tabular}{r@{\hspace{0.03in}}l}\textbf{function } & #1\\\textbf{ where } & #2\textbf{ is }#3\end{tabular}}{\textbf{}}
\algblockdefx{WFuncb}{EndWFuncb}[3]{\textbf{function }#1\textbf{ where }#2\textbf{ is }#3}{\textbf{}}
\algtext*{EndWFuncb}\algblockdefx{WWFunc}{EndWWFunc}[5]{\begin{tabular}{l@{\hspace{0.03in}}r@{\hspace{0.03in}}l}\textbf{function }#1&\textbf{where}&#2\textbf{ is }#3\\&\textbf{and}&#4\textbf{ is }#5\end{tabular}}{\textbf{}}
\algblockdefx{WWWFunc}{EndWWWFunc}[7]{\begin{tabular}{l@{\hspace{0.03in}}r@{\hspace{0.03in}}l}\textbf{function }#1&\textbf{where}&#2\textbf{ is }#3\\&\textbf{and}&#4\textbf{ is }#5\\&\textbf{and}&#6\textbf{ is }#7\end{tabular}}{\textbf{}}
\algblockdefx{Funcs}{EndFuncs}[1]{\textbf{functions }#1}{\textbf{}}
\algblockdefx{Funcsb}{EndFuncsb}[1]{\textbf{functions }#1}{\textbf{}}
\algtext*{EndFuncsb}\algblockdefx{Funcb}{EndFuncb}[1]{\textbf{function }#1}{\textbf{}}
\algtext*{EndFuncb}\algnewcommand\lFor[2]{\State\textbf{for}\ #1\ \textbf{do}\ #2}
\algnewcommand\lIf[2]{\State\textbf{if}\ #1\ \textbf{then}\ #2}

\algnewcommand\algorithmicforeach{\textbf{for each}}
\algdef{S}[FOR]{ForEach}[1]{\algorithmicforeach\ #1\ \algorithmicdo}

\newcommand{\hide}[1]{}

\hypersetup{
  colorlinks=false,
  linkcolor=purple,
  citecolor=purple,
  urlcolor=purple,
  pdfauthor={},
  pdftitle={},
  pdfsubject={},
  pdfkeywords={}
}

\makeatletter
\renewenvironment{proof}[1][\proofname]{\par
\pushQED{\qed}\normalfont
  \topsep0pt \partopsep0pt \trivlist
  \item[\hskip\labelsep
        \itshape
    #1\@addpunct{.}]\ignorespaces
}{\endtrivlist\@endpefalse
}
\makeatother

\newtagform{blue}{\color{BlueViolet}(}{)}

\newtheorem{theorem}{Theorem}

\newtheorem{lemma}[theorem]{Lemma}

\usepackage{pdfpages}

\AtBeginDocument{\typeout{FLOATSPACE textfloatsep=\the\textfloatsep\space floatsep=\the\floatsep\space
    intextsep=\the\intextsep\space dbltextfloatsep=\the\dbltextfloatsep\space
    dblfloatsep=\the\dblfloatsep\space above=\the\abovecaptionskip\space below=\the\belowcaptionskip}\setlength{\textfloatsep}{9pt plus 2pt minus 3pt}\setlength{\floatsep}{8pt plus 2pt minus 2pt}\setlength{\intextsep}{8pt plus 2pt minus 2pt}\setlength{\dbltextfloatsep}{9pt plus 2pt minus 3pt}\setlength{\dblfloatsep}{8pt plus 2pt minus 2pt}\setlength{\abovecaptionskip}{4pt}\setlength{\belowcaptionskip}{0pt}}

\begin{document}

\def\sectionautorefname{\S}
\def\subsectionautorefname{\S}
\def\figureautorefname{Fig.}
\def\tableautorefname{Tab.}
\def\algorithmautorefname{Alg.}
\def\equationautorefname{Eq.}

\title{Pearl: TEE-Enabled Fair Ordering with Minimal Replication}

\author{\IEEEauthorblockN{Vian Robotin, J\'er\'emie Decouchant}
\IEEEauthorblockA{Delft University of Technology, The Netherlands}
\and
\IEEEauthorblockN{Shaokang Xie\IEEEauthorrefmark{2}, Dakai Kang\IEEEauthorrefmark{2}\IEEEauthorrefmark{3}, Mohammad Sadoghi\IEEEauthorrefmark{2}}
\IEEEauthorblockA{\IEEEauthorrefmark{2}Exploratory Systems Lab, UC Davis, USA\\\IEEEauthorrefmark{3}Sei Labs}
}

\maketitle

\begin{abstract}
Recent research has investigated fair transaction ordering in blockchain consensus. An orthogonal line of work has demonstrated that equipping replicas with trusted execution environments (TEEs) enables more efficient consensus protocols with only $2f{+}1$ replicas.
In this work, we bridge these two areas and demonstrate that two key fair-ordering properties, ordering linearizability and $\gamma$-batch-order fairness with minimal batching, cannot be implemented in partially synchronous networks with $2f{+}1$ replicas, even when the replicas are equipped with TEEs.
We show that Byzantine replicas can circumvent TEE protections by manipulating the timing of transaction submissions to the enclave,
thereby biasing the resulting ordering indicators. However,
we prove that assuming a synchronous network allows both properties to be implemented with a reduced replica count.
We propose two mechanisms: one uses absolute TEE-attested timestamps to guarantee ordering linearizability in a system of $n \ge 2f{+}1$ replicas,
and the other uses TEE-maintained relative sequence numbers to guarantee $\gamma$-batch-order fairness when $n > \frac{2\gamma f}{2\gamma-1}$.
Ultimately, we integrate these mechanisms into a synchronous lane-based consensus algorithm, achieving strong fair-ordering properties and superior performance while requiring fewer replicas than representative state-of-the-art baselines.
\end{abstract}

\noindent\textbf{Artifact availability:} \url{https://github.com/VianRobotin/incubator-resilientdb/tree/vian}

\section{Introduction}
\label{sec:intro}

Distributed ledgers rely on a Byzantine fault-tolerant (BFT) consensus service to produce a total order over client transactions~\cite{nakamoto2008bitcoin,wood2014ethereum}.
Yet classical leader-based and DAG-based BFT consensus protocols
\cite{castro1999pbft,yin2019hotstuff,keidar2021all,danezis2022narwhal,spiegelman2022bullshark,autobahn} only require agreement on a total order, leaving the position of each transaction entirely to (potentially malicious) protocol participants.
In financial applications, this freedom has been weaponized:
the maximal extractable value (MEV) literature has documented the extraction of hundreds of millions of dollars per year from end users through transaction reordering attacks~\cite{daian2019flash,eskandari2020sok,heimbach2022sok,malkhi2022maximal,weintraub2022flash}.

A growing body of work formalizes what a ``fair'' total order should look like.
Receive-order fairness, defined in Aequitas~\cite{kelkar2020order}, promises that if all correct replicas observe a transaction $t_1$ before a transaction $t_2$, then $t_1$ is committed before $t_2$. This property, however, is impossible to achieve in fully asynchronous settings because of Condorcet cycles~\cite{gehrlein1983condorcet,vafadar2023condorcet}.
$\gamma$-batch-order fairness~\cite{kelkar2023themis} sidesteps the impossibility by relaxing the property and grouping transactions from Condorcet cycles into the same batch.
Ordering linearizability~\cite{zhang2020byzantine} follows a different approach by attaching a numerical ordering indicator (typically a timestamp) to each transaction and requiring the system to order transactions using the median of $2f{+}1$ ordering indicators per transaction.
Subsequent work has explored attacks against fair ordering protocols~\cite{vafadar2023condorcet,park2025frontrunning}, faster variants of $\gamma$-batch-order fairness~\cite{cachin2022quick}, leaderless variants of fair ordering protocols on DAGs~\cite{raikwar2023fairness,kang2026fairdagconsensusfairnessmultiproposer,segalini2026tilikum,putnik2026herring}, and median- or threshold-encryption-based defenses~\cite{wendy2020kursawe,wendy2021kursawe,alpos2023eating,zarbafian2023lyra,gramoli2024aoab}.
To the best of our knowledge, all fair ordering protocols proposed to date require at least $3f{+}1$ replicas for ordering linearizability, and between $3f{+}1$ and $4f{+}1$ replicas for $\gamma$-batch-order fairness.

Another line of work equipped each replica with a trusted execution environment (TEE), a hardware-isolated enclave that can run authenticated code and produce unforgeable attestations, to lower the replication threshold and improve the performance of BFT consensus from $3f{+}1$ to $2f{+}1$ by ruling out equivocation~\cite{veronese2013minbft,behl2017hybster, decouchant2022damysus,rahli2024oneshot}.
It is therefore natural to ask whether TEEs can enable fair ordering. Prior work has explored this direction but still assumes $3f{+}1$ replicas~\cite{stathakopoulou2021adding,zarbafian2023aion}. Using TEEs to lower the replica count while maintaining fair ordering is not straightforward because TEEs cannot directly read or send network messages: a Byzantine host can control when a transaction is shown to its enclave.

In this work, we make the following \textbf{contributions}:

$\bullet$ We prove that ordering linearizability and $\gamma$-batch-order fairness cannot be implemented in partially synchronous networks with only $2f{+}1$ replicas, even with TEEs (\S\ref{sec:impossibility}).

$\bullet$ We then show that this impossibility can be overcome in synchronous networks (\S\ref{sec:impossibility}). We propose two mechanisms: one achieving ordering linearizability with absolute timestamps, requiring $n \ge 2f{+}1$ replicas, and another achieving $\gamma$-batch-order fairness with relative local orderings, requiring $n > \tfrac{2\gamma f}{2\gamma - 1}$. Both improve on the state of the art.

$\bullet$ We compose these mechanisms with lane-based dissemination and synchronous consensus to obtain the \pname{} protocols: a shared core (\S\ref{sec:common-core}), \nameol{} for ordering linearizability (\S\ref{sec:impl-ol}), and \namebof{} for $\gamma$-batch-order fairness (\S\ref{sec:impl-bof}). \pname{} builds on modified versions of Autobahn~\cite{autobahn} and Sync~HotStuff~\cite{synchotstuff}, respectively. Table~\ref{tab:fair-ordering-comparison} compares both variants with the related work. 

\lean{$\bullet$ We machine-checked the impossibility results and the safety and liveness of both variants in Lean~4, on a model of the protocols.}

$\bullet$ We deploy \pname{} on the DAS-5 cluster~\cite{bal2016medium} and compare it with four fair-ordering baselines (Pomp\=e~\cite{zhang2020byzantine}, Themis~\cite{kelkar2023themis}, FairDAG-AB~\cite{kang2026fairdagconsensusfairnessmultiproposer}, and FairDAG-RL~\cite{kang2026fairdagconsensusfairnessmultiproposer}) and with Tusk~\cite{danezis2022narwhal} as a vanilla DAG-BFT reference. \rev{\pname{} achieves $3$--$13\times$ the throughput of the strongest fair-ordering baseline, while also providing lower latency ($230$--$300$~ms).}

\section{System Model and Objectives}

\textbf{System Model.} We consider a system of $n$ replicas running a consensus service for a blockchain. Up to $f$ replicas may be Byzantine. Clients submit their transactions to every replica to have them committed in the chain. We make no synchrony assumption on the links between clients and replicas. A transaction only needs to reach one correct replica, and our fairness guarantees are stated from that point on.
Our impossibility results assume a partially synchronous inter-replica network.\footnote{A network is partially synchronous if there exists a known bound $\Delta$ and an unknown time $\mathsf{GST}$ such that, after $\mathsf{GST}$, every message sent between correct replicas is delivered within $\Delta$ time~\cite{dwork1988consensus}.}
By contrast, our fair-ordering protocols, \nameol{} and \namebof{}, assume a synchronous network with a known maximum transmission delay $\Delta$. They respectively require $n \ge 2f{+}1$ and $n > \frac{2\gamma f}{2\gamma-1}$.
Each correct replica has a local clock that it uses to measure durations, and we assume that these clocks advance at the rate of real time. We do not assume synchronized clocks.

\textbf{Trusted Execution Environments.} Each replica runs a trusted execution environment (TEE) capable of executing authenticated code and producing unforgeable attestations.
TEEs cannot directly communicate over the network and rely on their (potentially Byzantine) host for communication. We assume remote authentication and access to signature and encryption schemes, but exclude rollback and side-channel attacks~\cite{rahli2026pallas}.
TEEs have no trusted clock but can maintain state that their host cannot read or tamper with. When a TEE needs a timestamp, it reads one from its host and enforces that successive attested values strictly increase. \lean{Through remote attestation, the enclaves of different replicas can also establish a shared key that no host learns.}
\nameol{} uses host-provided timestamps attested by TEEs, whereas \namebof{} uses TEE-maintained counters.

\textbf{Fair Ordering.} We aim to implement ordering linearizability~\cite{zhang2020byzantine} and $\gamma$-batch-order fairness~\cite{kelkar2020order,kelkar2023themis} in systems with $2f{+}1$ replicas, achieving this combination for the first time.

\begin{definition}[Ordering Linearizability]
\label{def:ol}
    For any two transactions $t_1$ and $t_2$,
    \rev{if every timestamp that a correct replica assigns to $t_1$ is smaller than every timestamp that a correct replica assigns to $t_2$, then $t_1$ is ordered before $t_2$.}
\end{definition}

\rev{The timestamp that a replica assigns to a transaction is the reading of its local clock when it observes the transaction\lean{, raised just above its previous timestamp if the two readings coincide}. Note, however, that ordering linearizability does not require clock synchronization.} 
Pompē~\cite{zhang2020byzantine} implements ordering linearizability using $3f{+}1$ non-TEE replicas by assigning each transaction a final ordering indicator equal to the median of $2f{+}1$ replica timestamps.
This guarantees that the final timestamp is both lower- and upper-bounded by timestamps from correct replicas.
Differently, $\gamma$-batch-order fairness~\cite{kelkar2020order,kelkar2023themis} relies on replicas' relative transaction orders rather than absolute timestamps.

\begin{definition}[$\gamma$-batch-order fairness with minimal batching]
\label{def:bof}
    For any two transactions $t_1$ and $t_2$,
    if at least a fraction $\gamma \in (\frac{1}{2}, 1]$ of all correct replicas receive $t_1$ before $t_2$ (denoted $t_1 \locprec t_2$), then $t_1$ must be ordered \add{before $t_2$ or in the same batch}.
    \rev{Transactions are placed in the same batch only if their preferences form a directed cycle in which each edge $t_1 \to t_2$ is supported by at least $q$ correct replicas that receive $t_1$ before $t_2$. The protocol fixes the \emph{batching support} $q \in \{1, \cdots, \lceil \gamma(n{-}f) \rceil\}$, ensuring batch minimality at support~$q$.}
    \add{A batch may be output over several execution steps and need not be contiguous.}
\end{definition}

We adopt the formulation introduced by Themis~\cite{kelkar2023themis}, which places transactions in the same batch only when necessary to resolve cyclic ordering dependencies.
\add{Condorcet cycles can chain without bound, so a protocol that waits for a complete batch is not live.}

\section{Impossibility of Fair Ordering with $n=2f{+}1$ in Partially-Synchronous Networks}
\label{sec:impossibility}

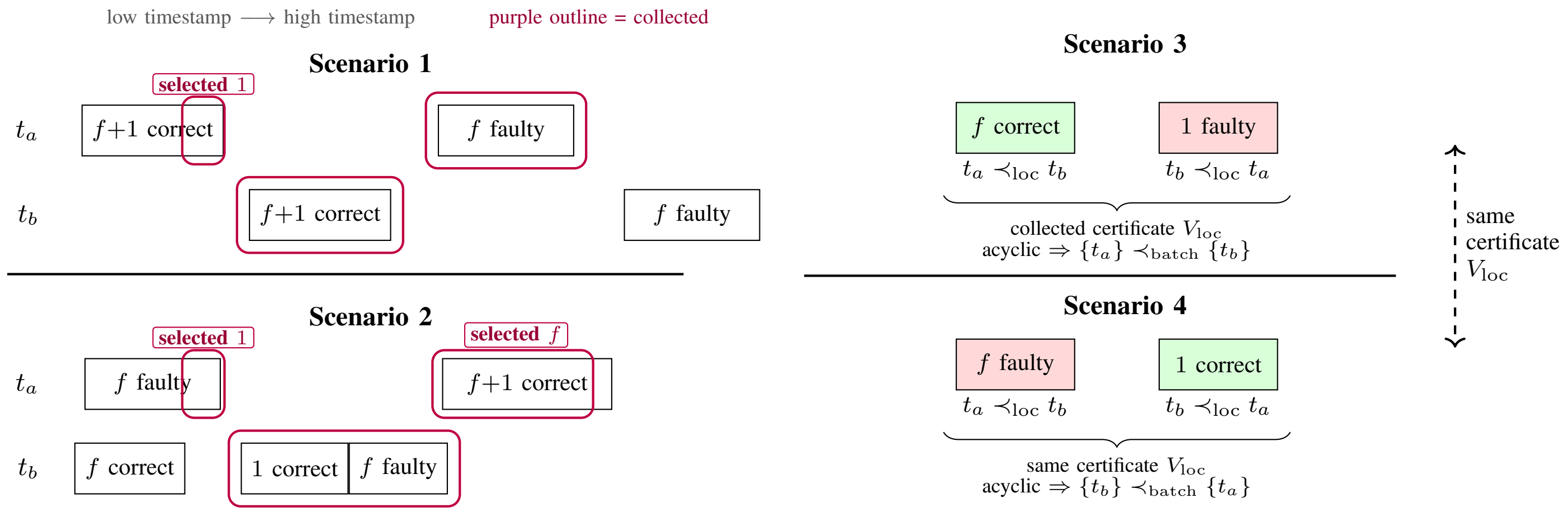
\begin{figure*}[t]
\centering
\begin{subfigure}[t]{0.48\textwidth}
\centering
\begin{tikzpicture}[
    box/.style={draw, rectangle, minimum height=0.6cm, minimum width=1.6cm,
        align=center, font=\footnotesize},
    circ/.style={draw=purple, thick, rounded corners, inner sep=4pt},
    countlabel/.style={draw=purple, rounded corners=1pt,
        font=\scriptsize\bfseries, text=purple!80!black,
        fill=white, inner xsep=2pt, inner ysep=1pt},
    txlabel/.style={anchor=east, font=\small},
    scenariolabel/.style={font=\bfseries\small}
]
\def\xA{0}
\def\xB{2.2}
\def\xC{4.4}
\def\xD{6.6}

\node[font=\scriptsize, text=black!65] at (1.5,3.72)
    {low timestamp $\longrightarrow$ high timestamp};
\node[font=\scriptsize, text=purple!80!black] at (5.5,3.72)
    {purple outline = collected};

\node[scenariolabel] at (2.8,3.2) {Scenario 1}; \node[txlabel] at (-1.0,2.4) {$t_a$};
\node[txlabel] at (-1.0,1.4) {$t_b$};
\node[box] (a1) at (\xA + 0.22,2.4) {$f{+}1$ correct};
\node[box] (a2) at (\xC,2.4) {$f$ faulty};
\node[box] (b1) at (\xB,1.4) {$f{+}1$ correct};
\node[box] (b2) at (\xD,1.4) {$f$ faulty};
\node[circ, fit=(a2)] {};
\node[circ, fit=(b1)] {};
\node[circ, minimum width=0.5cm, minimum height=0.8cm]
    at ([xshift=0.6cm]a1.center) {};
\node[countlabel] at ([xshift=0.6cm,yshift=0.55cm]a1.center)
    {selected $1$};

\draw[thick] (-1.5,0.7) -- (6.5,0.7);

\node[scenariolabel] at (2.8,0.2) {Scenario 2}; \node[txlabel] at (-1.0,-0.6) {$t_a$};
\node[txlabel] at (-1.0,-1.6) {$t_b$};
\node[box] (c1) at (\xA + 0.22, -0.6) {$f$ faulty};
\node[box, minimum width=2cm] (c2) at (\xC+0.25,-0.6) {$f{+}1$ correct};
\node[box, minimum width=1.3cm] (d0) at (\xA - 0.05,-1.6) {$f$ correct};
\node[box, minimum width=1.15cm] (d1) at (\xB - 0.3,-1.6) {$1$ correct};
\node[box, minimum width=1.15cm, anchor=west] (d2) at (d1.east) {$f$ faulty};
\node[circ, minimum width=1.9cm, minimum height=0.8cm]
    at ([xshift=-0.17cm]c2.center) {};
\node[circ, fit=(d1)(d2)] {};
\node[circ, minimum width=0.5cm, minimum height=0.8cm]
    at ([xshift=0.6cm]c1.center) {};
\node[countlabel] at ([xshift=0.6cm,yshift=0.55cm]c1.center)
    {selected $1$};
\node[countlabel] at ([xshift=-0.13cm,yshift=0.58cm]c2.center)
    {selected $f$};
\end{tikzpicture}
\caption{Ordering linearizability scenarios. Both scenarios produce the same $f{+}1$ absolute timestamps per transaction\lean{, signed by the same replicas}. Purple outlines mark the timestamps used per transaction.}
\label{fig:ol-impossibility}
\end{subfigure}
\hfill
\begin{subfigure}[t]{0.48\textwidth}
\centering
\begin{tikzpicture}[
    box/.style={draw, rectangle, minimum height=0.6cm, minimum width=1.4cm,
        align=center, font=\footnotesize},
    circ/.style={draw=purple, thick, rounded corners, inner sep=4pt},
    txlabel/.style={anchor=east, font=\small},
    scenariolabel/.style={font=\bfseries\small},
    brack/.style={decorate, decoration={brace, amplitude=5pt, mirror}}
]

\def\offset{1}

\node[scenariolabel] at (2.3,3.2)
    {Scenario 3}; 

\node[box, fill=green!15] (s1c) at (0 + \offset,2.2) {$f$ correct};
\node[font=\footnotesize] at (0 +\offset,1.7) {$t_a \locprec t_b$};

\node[box, fill=red!15] (s1f1) at (2.4 + \offset,2.2) {$1$ faulty};
\node[font=\footnotesize] at (2.4 + \offset,1.7) {$t_b \locprec t_a$};

\draw[brack] (-0.85 + \offset,1.4) -- (3.25 + \offset,1.4)
    node[midway, below=5pt, font=\scriptsize, align=center]
    {collected certificate $V_{\rm loc}$\\[-1pt]
     acyclic $\Rightarrow \{t_a\}\prec_{\rm batch}\{t_b\}$};

\draw[thick] (-1.5,0.45) -- (5.5,0.45);

\node[scenariolabel] at (2.3,0.1)
    {Scenario 4}; 

\node[box, fill=red!15] (s2f) at (0 + \offset,-0.6) {$f$ faulty};
\node[font=\footnotesize] at (0 + \offset,-1.1) {$t_a \locprec t_b$};

\node[box, fill=green!15] (s2c) at (2.4 + \offset,-0.6) {$1$ correct};
\node[font=\footnotesize] at (2.4 + \offset,-1.1) {$t_b \locprec t_a$};

\draw[brack] (-0.85 + \offset,-1.4) -- (3.25 + \offset,-1.4)
    node[midway, below=5pt, font=\scriptsize, align=center]
    {same certificate $V_{\rm loc}$\\[-1pt]
     acyclic $\Rightarrow \{t_b\}\prec_{\rm batch}\{t_a\}$};

\draw[<->, thick, dashed] (4.2 + \offset + 1,2.0) -- (4.2 + \offset + 1,-0.4)
    node[midway, right, font=\footnotesize, align=left]
    {same\\certificate\\$V_{\rm loc}$};

\end{tikzpicture}
\caption{$\gamma$-batch-order fairness scenarios.
Both scenarios produce the same $f{+}1$ local-order certificates, where a replica that observes a transaction $t_a$ before a transaction $t_b$ expresses it as $t_a \locprec t_b$.
}
\label{fig:bof-impossibility}
\end{subfigure}
\caption{Fair ordering impossibility with $2f{+}1$ replicas in partially synchronous networks.
}
\Description{Two side-by-side panels show indistinguishable executions for ordering linearizability and minimal $\gamma$-batch-order fairness under $2f{+}1$ replicas.}
\label{fig:ordering-impossibility}
\end{figure*}

In this section, we assume that the inter-replica network is partially synchronous.
We aim to design fair-ordering protocols that operate with $n \geq 2f{+}1$ replicas, which is the minimum number of replicas required to implement consensus.
In this setting, a consensus algorithm should order transactions based on $f{+}1$ ordering indicators per transaction.
Under this constraint, we show that neither ordering linearizability (based on absolute timestamps) nor $\gamma$-batch-order fairness (based on local relative orderings) can be guaranteed, even with TEEs.

\begin{lemma}
\label{lemma:ol}
    In a partially synchronous system of $n~=~2f{+}1$ replicas with up to $f$\lean{${}\geq 1$} Byzantine faults,
    no live ordering rule that decides from only $f{+}1$ signed absolute-timestamp reports per transaction can guarantee ordering linearizability.
\end{lemma}

\begin{proof}
We consider two scenarios that a protocol collecting $f{+}1$ absolute timestamps per transaction cannot distinguish, yet that require opposite ordering decisions.

    \emph{Scenario~1 (Fig.~\ref{fig:ordering-impossibility}).} Every correct timestamp of $t_a$ is smaller than every correct timestamp of $t_b$.
    Ordering linearizability therefore requires that $t_a$ be ordered before $t_b$.
    Meanwhile, the $f$ faulty replicas issue timestamps for $t_a$ that exceed all correct timestamps of $t_b$ and timestamps for $t_b$ that exceed all faulty timestamps of $t_a$.
    The adversary then arranges for the protocol to collect, for each transaction,
    the set of $f{+}1$ highlighted timestamps:
    one correct timestamp and $f$ faulty timestamps for $t_a$ and $f{+}1$ correct timestamps for $t_b$.
    From these collected timestamps, the protocol must, by assumption,
    decide $t_a \prec t_b$ to remain consistent with ordering linearizability.

    \emph{Scenario~2 (Fig.~\ref{fig:ordering-impossibility}).} \rev{Every correct timestamp of $t_b$ is smaller than every correct timestamp of $t_a$.}
    Ordering linearizability therefore requires that $t_b$ be ordered before $t_a$.
    The faulty replicas issue timestamps for $t_a$ that are smaller than all correct timestamps of $t_b$ and timestamps for $t_b$ that are larger than all correct timestamps of $t_a$.
    The adversary arranges for the protocol to collect $f$ correct timestamps and one faulty timestamp for $t_a$,
    and $f{+}1$ correct timestamps for $t_b$.
    Assuming that the collected timestamps are identical to those collected in Scenario~1, the protocol must make the same decision and order $t_a$ before $t_b$. However, in this case, this decision contradicts ordering linearizability, completing the proof.
    
    \qed{}
\end{proof}

\begin{lemma}
    \label{lemma:bof}
    In a partially synchronous system of $n~=~2f{+}1$ replicas with up to $f$\lean{${}\geq 1$} Byzantine faults,
    no live ordering rule that decides from $f{+}1$ signed local-order reports guarantees $\gamma$-batch-order fairness with minimal batching, for any $\gamma \in (\frac{1}{2}, 1]$.
\end{lemma}

\begin{proof}
    In the considered setting, a consensus protocol must order any pair of transactions using only $f{+}1$ local orderings that contain both $t_a$ and $t_b$, with $\gamma \in (\tfrac{1}{2}, 1]$.
    We exhibit two scenarios that cannot be distinguished and that require opposite ordering decisions according to $\gamma$-batch-order fairness. \rev{In both, all correct replicas observe a transaction before the other, so batch minimality prevents the two transactions from being batched together.}

    \emph{Scenario~3 (Fig.~\ref{fig:ordering-impossibility}).} All $f{+}1$ correct replicas receive $t_a$ before $t_b$. Thus, by $\gamma$-batch-order fairness, the protocol must order $t_a$ no later than $t_b$.
    The adversary arranges for the collected $f{+}1$ local orderings to consist of $f$ correct reports where $t_a \locprec t_b$ (i.e., $t_a$ appears before $t_b$) and one faulty report where $t_b \locprec t_a$. By assumption, the protocol must decide $t_a \prec t_b$.

    \emph{Scenario~4 (Fig.~\ref{fig:ordering-impossibility}).} All $f{+}1$ correct replicas receive $t_b$ before $t_a$.
The adversary now arranges for the collected $f{+}1$ local orderings to consist of $f$ faulty reports of $t_a \locprec t_b$ and one correct report of $t_b \locprec t_a$.
    The local relative orderings collected are indistinguishable from those in Scenario~3, so the protocol must make the same decision and order $t_a$ before $t_b$, which contradicts $\gamma$-batch-order fairness.
    \qed{}
\end{proof}

\begin{table*}[t]
\centering
\caption{Comparison of fair-ordering protocols.}
\vspace{-1mm}
\label{tab:fair-ordering-comparison}

\resizebox{\textwidth}{!}{\begin{tabular}{@{}p{4.6cm}llllcl@{}}
\toprule
\textbf{Protocol(s)} &
\textbf{Fairness} &
\textbf{Network} &
\textbf{Replicas} &
\textbf{TEE} &
\textbf{Liveness} &
\textbf{Technique} \\
\midrule

Pomp\=e~\cite{zhang2020byzantine},
AOAB~\cite{gramoli2024aoab}
& Ordering linearizability
& Partial synchrony
& $n \geq 3f+1$
& $\times$
& Standard
& Timestamp aggregation \\

Sync. Aequitas~\cite{kelkar2020order}
& $\gamma$-batch-order fairness
& Synchrony
& $n > \frac{2f}{2\gamma-1}$
& $\times$
& Weak
& Ordering/dependency graph \\

Async. Aequitas~\cite{kelkar2020order}
& $\gamma$-batch-order fairness
& Asynchrony
& $n > \frac{2f(\gamma+1)}{2\gamma-1}$
& $\times$
& Weak
& Ordering/dependency graph \\

Auncel~\cite{chen2024auncel}
& $\gamma$-batch-order fairness
& Partial synchrony
& $n > \frac{2f(\gamma+1)}{2\gamma-1}$
& $\times$
& Standard
& Ordering/dependency graph \\

Rashnu~\cite{nagda2024rashnu}
& Data-dependent order fairness
& Partial synchrony
& $n \geq 3f+1$
& $\times$
& Standard
& Data-dependent ordering graph \\

AuTIG~\cite{ren2025autig}
& $\gamma$-batch-order fairness
& Partial synchrony
& $n \geq 3f+1$
& $\times$
& Standard
& Incremental ordering graph + proofs \\

Themis~\cite{kelkar2023themis}
& $\gamma$-batch-order fairness
& Partial synchrony
& $n > \frac{2f(\gamma+1)}{2\gamma-1}$
& $\times$
& Standard
& Relative order + batching \\

FairDAG-AB~\cite{kang2026fairdagconsensusfairnessmultiproposer},
Tilikum-OL~\cite{segalini2026tilikum}
& Ordering linearizability
& Partial synchrony
& $n \geq 3f+1$
& $\times$
& Standard
& DAG-based ordering \\

FairDAG-RL~\cite{kang2026fairdagconsensusfairnessmultiproposer},
Tilikum-BOF~\cite{segalini2026tilikum}
& $\gamma$-batch-order fairness
& Partial synchrony
& $n > \frac{(2\gamma+1)f}{2\gamma-1}$
& $\times$
& Standard
& DAG-based ordering \\

Herring~\cite{putnik2026herring}
& $\gamma$-batch-order fairness
& Partial synchrony
& $n > \frac{4f}{2\gamma-1}$
& $\times$
& Standard
& DAG-based ordering \\

Fairy~\cite{stathakopoulou2021adding},
Aion~\cite{zarbafian2023aion}
& Input/time-based
& Partial synchrony
& $n \geq 3f+1$
& $\checkmark$
& Standard
& TEE-attested timestamps \\

\midrule

\textbf{\nameol{} (this work)}
& \textbf{Ordering linearizability}
& \textbf{Synchrony}
& $\mathbf{n \geq 2f+1}$
& $\mathbf{\checkmark}$
& \textbf{Standard}
& \textbf{TEE-attested host timestamps} \\

\textbf{\namebof{} (this work)}
& \textbf{$\gamma$-batch-order fairness}
& \textbf{Synchrony}
& $\mathbf{n > \frac{2\gamma f}{2\gamma-1}}$
& $\mathbf{\checkmark}$
& \textbf{Standard}
& \textbf{TEE-attested local orderings} \\

\bottomrule
\end{tabular}}
\end{table*}

\section{Fair Ordering Mechanisms in Synchronous Networks}
\label{sec:sync-fair}

We now state and prove the mechanisms that implement ordering linearizability and $\gamma$-batch-order fairness with fewer replicas in a synchronous network.

\begin{lemma}
\label{lemma:sync-ol}
    Let $C$ denote the set of correct replicas, and let $T_c(t)$ be the absolute timestamp that correct replica $c \in C$ produces for transaction~$t$.
    A system of $n = 2f{+}1$ replicas implements ordering linearizability if it assigns every transaction~$t$ a final ordering key $K(t)$ that falls within the range of its correct timestamps, i.e.,
    \(
        \min_{c \in C} T_c(t) \le K(t) \le \max_{c \in C} T_c(t),
    \)
    and orders transactions by increasing key.
\end{lemma}

\begin{proof}
    \rev{Assume that} every timestamp that a correct replica produces for $t_a$ is smaller than every timestamp that a correct replica produces for $t_b$, i.e., $\max_{c \in C} T_c(t_a) < \min_{c \in C} T_c(t_b)$.
    Since both keys fall within the range of their correct timestamps,
    \(
        K(t_a) \le \max_{c \in C} T_c(t_a) \\< \min_{c \in C} T_c(t_b) \le K(t_b),
    \)
    and the system, which orders transactions by increasing key, orders $t_a$ before $t_b$ as required by ordering linearizability. \qed
\end{proof}

\begin{lemma}
\label{sync-bof}
    Let $\gamma \in (\frac{1}{2}, 1]$, and let every replica report, for each pair of transactions, the order in which it received them.
    In a synchronous network of $n > \frac{2\gamma f}{2\gamma-1}$ replicas, consider the rule that orders $t_a$ no later than $t_b$ whenever at least $\gamma(n{-}f)$ of the counted reports state that $t_a$ was received first, where the counted reports include those of all $n{-}f$ correct replicas. This rule applies to every pair $(t_a, t_b)$ that Definition~\ref{def:bof} constrains, and then does not apply to $(t_b, t_a)$.
\end{lemma}

\begin{proof}
    Assume that $n > \tfrac{2\gamma f}{2\gamma-1}$ and consider transactions $t_a$ and $t_b$ such that at least $\gamma(n-f)$ correct replicas receive $t_a$ before $t_b$, so that Definition~\ref{def:bof} requires $t_a$ to be ordered no later than~$t_b$.
    Because the counted reports include those of every correct replica, these $\gamma(n{-}f)$ replicas all report $t_a \locprec t_b$, so the rule applies to $(t_a, t_b)$. It remains to show that it does not also apply to $(t_b, t_a)$.
    \lean{Each replica reports one order for the pair, so there are at most $n - \gamma(n-f)$ reports that support $t_b \locprec t_a$. It is not possible to have $\gamma(n-f)$ such reports, because we would then have 
    \(
    2\gamma (n - f) \leq n,
    \)}
    which rearranges to $(2\gamma - 1)n \leq 2\gamma f$, that is, $n \leq \tfrac{2\gamma f}{2\gamma - 1}$, which contradicts our assumption.
    The rule therefore applies to $(t_a, t_b)$ only. Pairs to which it applies along a cycle are placed in the same batch (\S\ref{sec:bof-ordering}). \qed{}
\end{proof}

\section{\pname{}: Common Core}
\label{sec:common-core}

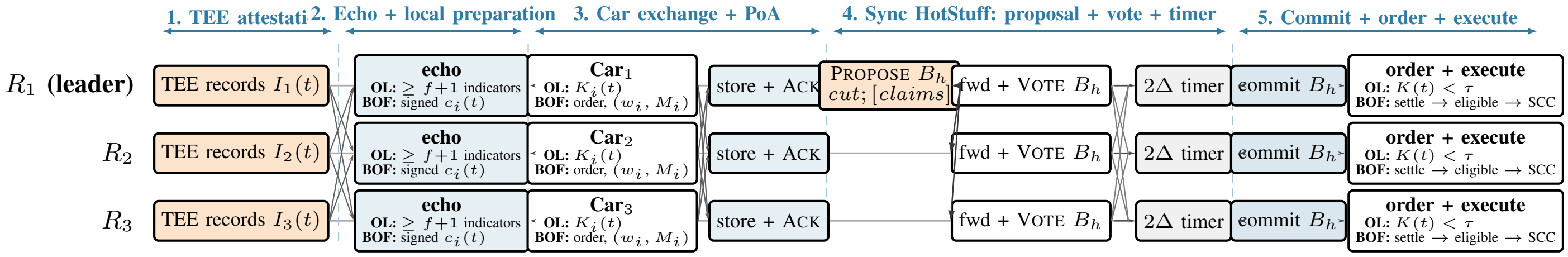
\begin{figure*}[!t]
\centering
\resizebox{\textwidth}{!}{
    \begin{tikzpicture}[
        >=Latex,
        thick,
        font=\small,
        lane/.style={draw=gray!65, line width=0.55pt},
        stage/.style={<->, >={Latex[length=1.25mm,width=0.85mm]},
                      draw=highlight2, line width=0.8pt},
        phaseboundary/.style={draw=highlight2!45, densely dashed,
                              line width=0.42pt},
        stagelabel/.style={font=\scriptsize\bfseries, text=highlight2,
                          fill=white, inner sep=1pt, align=center},
        state/.style={draw, fill=white, rounded corners=1.5pt,
                      minimum height=0.48cm, inner xsep=2.5pt, inner ysep=1.5pt,
                      align=center, font=\scriptsize},
        tee/.style={state, fill=highlight1!25},
        local/.style={state, fill=highlight2!12},
        leader/.style={state, fill=highlight1!25},
        timer/.style={state, fill=gray!12},
        commit/.style={state, fill=highlight2!18},
        msg/.style={-{Latex[length=1.25mm,width=0.85mm]},
                    line width=0.48pt, draw=black!78},
        xmsg/.style={-{Stealth[length=0.78mm,width=0.50mm]},
                     line width=0.42pt, draw=black!62,
                     shorten >=0.5pt},
        fanmsg/.style={line width=0.42pt, draw=black!52},
        localstep/.style={-{Stealth[length=0.72mm,width=0.46mm]},
                          line width=0.45pt, draw=black!72,
                          shorten <=1pt, shorten >=1.25pt},
        flowstep/.style={-{Stealth[length=0.90mm,width=0.58mm]},
                         line width=0.48pt, draw=black!76,
                         shorten <=2pt, shorten >=3.2pt},
    ]
    
\useasboundingbox (-1.45,0.98) rectangle (16.88,-1.96);
    
\def\variantnode#1#2#3{\begingroup
        \tiny
        \renewcommand{\arraystretch}{1}\begin{tabular}{@{}r@{\;}l@{}}
            \multicolumn{2}{c}{\scriptsize\bfseries\strut #1}\\[-1pt]
            \textbf{OL:}  & \strut #2\\[-1pt]
            \textbf{BOF:} & \strut #3
        \end{tabular}\endgroup}
    
\def\yone{0}
    \def\ytwo{-0.82}
    \def\ythree{-1.64}
    \foreach \y in {\yone,\ytwo,\ythree}
        {\draw[lane] (-0.05,\y) -- (16.78,\y);}
    
    \node[anchor=east, font=\small\bfseries] at (-0.15,\yone)
        {$R_1$ (leader)};
    \node[anchor=east, font=\small\bfseries] at (-0.15,\ytwo) {$R_2$};
    \node[anchor=east, font=\small\bfseries] at (-0.15,\ythree) {$R_3$};
    
\foreach \x in {2.20,4.50,8.12,13.04}
        {\draw[phaseboundary] (\x,0.51) -- (\x,-1.90);}
    
\draw[stage] (0.05,0.66) -- (2.14,0.66)
        node[midway, above=1pt, stagelabel] {1. TEE attestation};
    \draw[stage] (2.26,0.66) -- (4.44,0.66)
        node[midway, above=1pt, stagelabel]
        {2. Echo + local preparation};
    \draw[stage] (4.56,0.66) -- (8.06,0.66)
        node[midway, above=1pt, stagelabel]
        {3. Car exchange + PoA};
    \draw[stage] (8.18,0.66) -- (12.98,0.66)
        node[midway, above=1pt, stagelabel]
        {4. Sync HotStuff: proposal + vote + timer};
    \draw[stage] (13.10,0.66) -- (16.75,0.66)
        node[midway, above=1pt, stagelabel]
        {5. Commit + order + execute};
    
\foreach \i/\y in {1/\yone,2/\ytwo,3/\ythree}
        {\node[tee] (att\i) at (1.02,\y)
            {TEE records $I_{\i}(t)$};}
    
\foreach \i/\y in {1/\yone,2/\ytwo,3/\ythree}
        {\node[local] (echo\i) at (3.45,\y)
            {\variantnode{echo}{$\ge f{+}1$ indicators}{signed $c_i(t)$}};}
    
    \foreach \src/\dst in {att1/echo2,att1/echo3,
                             att2/echo1,att2/echo3,
                             att3/echo1,att3/echo2}
        {\draw[xmsg] (\src.east) -- (\dst.west);}
    
\foreach \i/\y in {1/\yone,2/\ytwo,3/\ythree}
        {\node[state] (car\i) at (5.52,\y)
            {\variantnode{Car$_{\i}$}{$K_i(t)$}{order, $(w_i,M_i)$}};
         \node[local] (ack\i) at (7.42,\y)
            {store + \textsc{Ack}};
         \draw[localstep] (echo\i.east) -- (car\i.west);}
    
    \foreach \src/\dst in {car1/ack2,car1/ack3,
                             car2/ack1,car2/ack3,
                             car3/ack1,car3/ack2}
        {\draw[xmsg] (\src.east) -- (\dst.west);}
    
\node[leader] (proposal) at (8.88,\yone)
        {\textsc{Propose} $B_h$\\[-1pt]$\mathit{cut};[\mathit{claims}]$};
    
    \foreach \i/\y in {1/\yone,2/\ytwo,3/\ythree}
        {\node[state] (vote\i) at (10.60,\y)
            {fwd + \textsc{Vote} $B_h$};
         \coordinate (recv\i) at (11.78,\y);
         \node[timer] (timer\i) at (12.45,\y)
            {$2\Delta$ timer};}
    
    \draw[msg] (proposal.east) -- (vote1.west);
    \draw[msg] (proposal.east) -- (vote2.west);
    \draw[msg] (proposal.east) -- (vote3.west);
    
\foreach \i in {1,2,3}
        {\draw[localstep, shorten >=0.15pt] (vote\i.east) -- (timer\i.west);}
    \foreach \src/\dst in {vote1/recv2,vote1/recv3,
                             vote2/recv1,vote2/recv3,
                             vote3/recv1,vote3/recv2}
        {\draw[fanmsg] (\src.east) -- (\dst);}
    
\foreach \i/\y in {1/\yone,2/\ytwo,3/\ythree}
        {\node[commit] (commit\i) at (13.72,\y) {commit $B_h$};
         \node[state] (exec\i) at (15.75,\y)
            {\variantnode{order + execute}{$K(t)<\tau$}{settle $\to$ eligible $\to$ SCC}};
         \draw[flowstep] (timer\i.east) -- (commit\i.west);
         \draw[localstep] (commit\i.east) -- (exec\i.west);}
    
    \end{tikzpicture}
}
\caption{Steady-state workflow of \pname{} (illustrated with $f=1$ and $n=3$); $R_1$ is the leader, and time flows from left to right. Here, $I_i=T_i$ in \nameol{} and $I_i=s_i$ in \namebof{}. Inline OL/BOF labels distinguish the variant-specific echo, Car-preparation, and post-commit ordering steps; Car dissemination and acknowledgment (forming a PoA) and the Sync HotStuff proposal/vote/timer path are shared. If its $2\Delta$ timer expires without equivocation or a view change, each replica commits locally without an additional commit-message round, derives the variant-specific order from committed data, and executes eligible transactions.}
\Description{Three horizontal replica lanes show five left-to-right phases: TEE attestation; transaction and indicator echo; variant-specific Car preparation followed by Car dissemination, acknowledgment, and PoA formation; Sync HotStuff proposal, forwarding, voting, and a local timer; and local commitment followed by variant-specific ordering and eligible execution.}
\label{fig:pearl-overview}
\end{figure*}

\pname{} turns the mechanisms of \S\ref{sec:sync-fair} into two
 variants that share the same two layers: a TEE-assisted dissemination layer (Alg.~\ref{alg:tee-replica}), and a synchronous consensus layer (Alg.~\ref{alg:consensus}). \rev{This design enables nodes to concurrently propose transaction blocks, while the synchronous consensus layer ensures the ordering mechanisms operate correctly.} The dissemination layer guarantees that transactions remain retrievable from at least one correct replica and associates each transaction with unforgeable ordering evidence. The consensus layer then commits transaction digests and variant-specific ordering evidence, allowing the execution layer to deterministically derive the final transaction order.
Fig.~\ref{fig:pearl-overview} sketches the resulting five-phase pipeline.

\rev{\textbf{Building blocks in a nutshell.} Autobahn~\cite{autobahn} is a partially synchronous protocol that separates data dissemination from consensus. Each replica disseminates batches of transactions, called Cars, in its own \emph{lane}, a hash chain that it extends at its own pace. A Car is certified available once $f{+}1$ replicas acknowledge it, so at least one correct replica holds it. Its consensus layer agrees on \emph{tip cuts}, i.e., one certified tip per lane, and committing a tip cut commits every Car in the prefix of each lane. Sync~HotStuff~\cite{synchotstuff} is a synchronous consensus protocol for $n \geq 2f{+}1$ replicas. A stable leader proposes a chain of blocks, each carrying a certificate of $f{+}1$ votes for its predecessor. Replicas forward every proposal and broadcast their votes, and a replica commits a block $2\Delta$ after voting for it if it has observed no equivocation in the meantime.}

The \pname{} variants, \nameol{} (\S\ref{sec:impl-ol}) and \namebof{} (\S\ref{sec:impl-bof}), differ in the authenticated evidence that their TEEs issue and in the rule that derives the final transaction order.

    \subsection{TEE-Attested Ordering Evidence}
    \label{sec:tee-indicators}

\textbf{Ordering evidence.} Correct clients send each of their transactions to every replica.
When a replica receives a transaction~$t$ for the first time, either directly from a client or from another replica, it forwards it to its local TEE to record the local order and generate authenticated evidence (Alg.~\ref{alg:tee-replica}, l.~\ref{ln:first-receive}).
The TEE keeps in its protected memory a set~$\mathcal{H}$ of the identifiers of the transactions it has already attested (Alg.~\ref{alg:tee-replica}, l.~\ref{ln:tee-history}),
where $\mathsf{id}(t)$ is the hash digest of~$t$.
If $\mathsf{id}(t) \in \mathcal{H}$, the TEE avoids processing transaction $t$ a second time and returns nothing (Alg.~\ref{alg:tee-replica}, l.~\ref{ln:dedup}).
Otherwise it inserts $\mathsf{id}(t)$ into~$\mathcal{H}$ (Alg.~\ref{alg:tee-replica}, l.~\ref{ln:insert-history}) and produces variant-specific evidence.
In \nameol{}, the enclave accepts a host-provided timestamp~$T$ only if it exceeds the last timestamp it attested, and returns $\sigma_{\mathrm{OL}}=\mathsf{Sign}_{\mathit{TEE}}(t,T)$ (Alg.~\ref{alg:tee-replica}, ll.~\ref{ln:ol-time}--\ref{ln:ol-sign}).
\lean{A correct host therefore submits the larger of its clock reading and a value just above the last attested timestamp~$T_{\mathit{last}}$.}
In \namebof{}, the enclave increments its local sequence number~$s$, appends~$t$ to its open Car~$c$, and returns the placement evidence $\sigma_{\mathrm{BOF}}=\mathsf{Sign}_{\mathit{TEE}}(t,c)$ (ll.~\ref{ln:bof-place}--\ref{ln:bof-sign}).
The signed Car later authenticates the resulting sequence order, while the placement signature separately binds~$t$ to the Car index announced during echoing.
Thus, each TEE's timestamp sequence in \nameol{} and transaction sequence in \namebof{} is unique and monotonically increasing, but the two variants do not share a common signed-indicator format.

\textbf{Non-equivocation.} Because $\mathcal{H}$ is enclave-resident and checked before each attestation (Alg.~\ref{alg:tee-replica}, l.~\ref{ln:dedup}), a Byzantine host cannot obtain multiple attestations for the same transaction or report different authenticated timestamps or placements for it to different replicas.
However, a faulty host retains several capabilities that a TEE cannot prevent (\S\ref{sec:impossibility}), because an enclave only observes the inputs forwarded to it by its host. Thus, the host may delay forwarding a transaction to the enclave or submit one of its transactions early, respectively inflating or deflating its indicator. Furthermore, it may withhold an attestation instead of broadcasting it. In \nameol{}, the host also chooses the timestamps fed to its enclave, subject only to the monotonicity check.

\textbf{Enclave state.}
\rev{The enclave keeps~$\mathcal{H}$ for its lifetime, because a host able to trigger the deletion of an identifier could obtain a second attestation for its transaction. For garbage collection, the enclave can delete an identifier after verifying a proof that its transaction was executed.}

    \subsection{Data Dissemination and Ordering-Evidence Collection}
    \label{sec:data-layer}

\textbf{Transaction dissemination.} \pname{} borrows the data dissemination layer of Autobahn~\cite{autobahn}, but adapts it to a lower number of replicas under synchrony. Unlike the DAG-BFT protocols it takes inspiration from~\cite{danezis2022narwhal,spiegelman2022bullshark}, Autobahn builds no graph of cross-referencing proposals: its data layer consists of $n$ independent \emph{lanes}, one per replica, each a hash chain that its owner extends at its own pace.
Replicas assemble the \rev{digests of the} transactions they have received into batches (called Cars in Autobahn, Alg.~\ref{alg:tee-replica}, ll.~\ref{ln:car-bof} and~\ref{ln:car-ol}), which are broadcast (in \textsc{Prop} messages, l.~\ref{ln:broadcast-car}) and acknowledged by the other replicas (in \textsc{Ack} messages\footnote{Autobahn calls them \textsc{Vote} messages, which we rename to avoid confusion with the consensus votes of \S\ref{sec:consensus-layer}.}, l.~\ref{ln:send-ack}).

A correct replica seals its open Car once it is full or once an interval~$X$ has elapsed since its previous seal, whichever comes first.
We set $X = \Delta$, the time a Car takes to reach every replica: at this cadence a replica has already received and acknowledged a Car's predecessor when the Car itself arrives, and can therefore acknowledge every Car on reception.
Cars are chained into per-replica lanes: a replica votes for the Car at position~$i$ of a lane only if it references the Car at position $i{-}1$, which the replica has already received and voted on.
In both variants the enclave controls the content of a Car: in \namebof{} it signs a Car only if the transactions appear in the order it observed them, while in \nameol{} it checks that it has attested all transactions and certified their local ordering keys (\S\ref{sec:ordering-logic}).
Each Car carries the $f{+}1$ signed acknowledgments of the most recent Car of the same lane for which its owner has assembled them, forming a Proof of Availability (PoA). The availability of a lane prefix still follows from the voting rule above.
\rev{A replica acknowledges a Car only once it holds the body of every transaction the Car references, fetching any missing body from the Car's owner and then from all replicas, so the $f{+}1$ acknowledgments of a PoA include a correct replica able to serve every body.}
Cars also contain protocol-specific ordering information for the transactions they contain (see \S\ref{sec:impl-ol}--\ref{sec:impl-bof}). Each Car finally carries a \emph{frontier}, which we define below.

\textbf{Transaction echoing and ordering-evidence collection.} Upon receiving a transaction $t$ for the first time, a replica produces variant-specific TEE evidence and retransmits $t$ to all other replicas in a dedicated message (Alg.~\ref{alg:tee-replica}, l.~\ref{ln:broadcast-indicator})\footnote{Note that this retransmission is also the known defense against the Ambush attack~\cite{park2025frontrunning}, which targets $\gamma$-batch-order fairness protocols.}. In \nameol{}, this message also carries the locally generated timestamp~$T$ and its TEE signature~$\sigma_{\mathrm{OL}}$.
In \namebof{}, it carries instead the \emph{placement announcement} $c_r(t)$, the index of the Car of lane~$r$ in which the enclave has just placed~$t$. This announcement is an unforgeable commitment on where~$t$ appears in $r$'s local order.
Every recipient verifies the complete variant-specific signed tuple before storing it (Alg.~\ref{alg:tee-replica}, ll.~\ref{ln:verify-ol-echo}--\ref{ln:store-bof-echo}).
A replica might receive a transaction for the first time either directly from a client, \rev{by fetching it for a Car it acknowledges}, or after this retransmission.
The first correct replica that receives a transaction retransmits it immediately, so every correct replica receives it within~$\Delta$ of that replica, and broadcasts its attestation upon reception. This holds whatever the delays between clients and replicas.
Consequently, at most $2\Delta$ after it first receives a transaction~$t$, a correct replica holds the attestation of every correct replica for~$t$, i.e., at least $f{+}1$ and up to~$n$ of them, which is why the collection window of l.~\ref{ln:collection-window} lasts $2\Delta$.
Both variants use this window, at different points of their pipeline. In \nameol{} it precedes placement: once it closes, the replica reduces $\mathit{timestamps}[t]$ to a \emph{local ordering key} $K_r(t)$ (Alg.~\ref{alg:tee-replica}, l.~\ref{ln:finalize}) and places~$t$ in its next Car (Alg.~\ref{alg:tee-replica}, l.~\ref{ln:car-ol}). In \namebof{} placement is immediate (Alg.~\ref{alg:tee-replica}, l.~\ref{ln:car-bof}), and the window instead governs the frontier defined next.

\textbf{Frontiers.} Sealing a Car fixes a frontier from which the protocol variants decide, from the committed Cars alone, when a transaction may be ordered.
\nameol{} seals the \emph{completion frontier} $G_r$ (l.~\ref{ln:frontier}), the largest indicator \lean{that the enclave has issued} such that every transaction $r$ attested at or below it has its local ordering key in that Car or in an earlier one. \lean{The enclave places each ordering key in a single Car. The frontier of a lane therefore never decreases from one Car to the next, since a transaction that is attested after a Car was sealed receives an ordering indicator that is greater than every earlier one.} The enclave derives $G_r$ from the set $\mathcal{H}$ of identifiers it has attested (l.~\ref{ln:tee-history}). A host may hold its frontier back, by queueing transactions before sealing a Car, but can never advance it past a transaction whose key it has withheld, nor past the largest indicator its enclave has issued (otherwise a transaction attested later could fall at or below an already-sealed frontier).
\namebof{} seals the \emph{reception frontier} $(w_r, M_r)$ (l.~\ref{ln:frontier-bof}). Its first component $w_r$ is the highest position of lane~$r$\lean{, no higher than the Car that carries it,} such that the collection window of every transaction placed at or below it has closed. Its second component gives, for every replica~$j$, the largest placement announcement $M_r[j]$ that $r$ collected from~$j$ for those transactions, carried with the enclave signature of~$j$ that certifies it. A correct~$r$ therefore seals $M_r[j] \geq c_j(t)$ for every correct~$j$ and every transaction~$t$ placed at or below~$w_r$: the window of~$t$ closed $2\Delta$ after $r$ received~$t$, by which time $j$'s announcement for~$t$ had arrived. A faulty host may hold $w_r$ back or omit entries from~$M_r$, but the signatures prevent it from reporting an index that $j$'s enclave never opened, and \S\ref{sec:bof-eligibility} tolerates under-reporting by aggregating the frontiers of $f{+}1$ lanes.
Because a frontier only advances when the enclave issues new indicators, correct replicas submit a no-op \emph{heartbeat} transaction to their enclave whenever no client transaction has arrived for a configurable period, so that frontiers keep advancing when client traffic pauses.

\begin{algorithm}[t]
\caption{\pname{}'s transaction dissemination and ordering-evidence collection at replica $r$. \nameol{} and \namebof{}-specific instructions are indicated using OL and BOF, respectively; routine lane-progression checks are omitted.}
\label{alg:tee-replica}
\algfont
\begin{algorithmic}[1]
\State \textbf{Global (OL):} $\mathit{timestamps}[t]$: signed timestamps, one per replica
\State \textbf{Global (BOF):} $\mathit{placements}[t]$: signed Car indices, one per replica
\State \textbf{Global:} $\mathit{lane} \gets \emptyset$
    \Comment{Local sequence of Cars}
\State \textbf{Global (inside TEE):} $\mathcal{H} \gets \emptyset$
    \Comment{Identifiers already attested}
    \label{ln:tee-history}

\Procedure{OnFirstReceive}{$t$}
    \label{ln:first-receive}
\If{$\mathsf{id}(t) \notin \mathcal{H}$}
        \Comment{TEE refuses a second attestation}
        \label{ln:dedup}
        \State $\mathcal{H} \gets \mathcal{H} \cup \{\mathsf{id}(t)\}$
            \label{ln:insert-history}
        \If{\textsc{OL}}
            \State $T \gets \lean{\max(}\mathsf{LocalClock}()\lean{,\, T_{\mathit{last}}{+}\varepsilon)}$
                \Comment{\rev{host-provided}}
                \label{ln:ol-time}
            \State $\sigma_{\mathrm{OL}} \gets \mathsf{TEE.Sign}(t,T)$
                \Comment{\rev{refused unless $T$ exceeds the last attested}}
                \label{ln:ol-sign}
            \State $\mathit{timestamps}[t].\mathsf{add}(T,\sigma_{\mathrm{OL}})$
            \State $\mathsf{BroadcastOLEcho}(t,T,\sigma_{\mathrm{OL}})$
                \label{ln:broadcast-indicator}
        \Else \Comment{\textsc{BOF}}
            \State $c \gets \mathsf{TEE.PlaceInOpenCar}(t)$
                \Comment{increment $s$ and append $t$ to Car $c$}
                \label{ln:bof-place}
                \label{ln:car-bof}
            \State $\sigma_{\mathrm{BOF}} \gets \mathsf{TEE.Sign}(t,c)$
                \label{ln:bof-sign}
            \State $\mathit{placements}[t].\mathsf{add}(c,\sigma_{\mathrm{BOF}})$
            \State $\mathsf{BroadcastBOFEcho}(t,c,\sigma_{\mathrm{BOF}})$
        \EndIf
    \EndIf
\EndProcedure

\Procedure{OnCarReady}{$\mathit{car}$} \Comment{append to lane}
    \State $\mathit{car}.G \gets \mathsf{TEE.CompletionFrontier}()$ \Comment{OL only}
          \label{ln:frontier}
    \State $\mathit{car}.(w,M) \gets \mathsf{TEE.ReceptionFrontier}()$
          \Comment{BOF only}
          \label{ln:frontier-bof}
    \State $\mathsf{Broadcast}(\textsc{Prop},\mathit{car})$ \Comment{with TEE attestation}
        \label{ln:broadcast-car}
\EndProcedure

\Procedure{OnReceiveProp}{$\mathit{car}, p$}
\State $\mathsf{FetchMissingBodies}(\mathit{car})$
        \Comment{\textsc{OnFirstReceive} on each}
        \label{ln:fetch-bodies}
    \State $\mathsf{StoreCar}(\mathit{car})$
        \label{ln:store-car}
    \State $\mathsf{Send}(p,\textsc{Ack},\mathit{car})$
        \label{ln:send-ack}
\EndProcedure

\Procedure{OnReceiveOLEcho}{$t,T,\sigma_{\mathrm{OL}}$}
    \State \Call{OnFirstReceive}{$t$}
        \Comment{no effect if already attested}
    \State \textbf{if} $\mathsf{TEE.Verify}(t,T,\sigma_{\mathrm{OL}})$ \textbf{then}
        \label{ln:verify-ol-echo}
        $\mathit{timestamps}[t].\mathsf{add}(T,\sigma_{\mathrm{OL}})$
            \label{ln:store-ol-echo}
\EndProcedure

\Procedure{OnReceiveBOFEcho}{$t,c,\sigma_{\mathrm{BOF}}$}
    \State \Call{OnFirstReceive}{$t$}
        \Comment{no effect if already attested}
    \State \textbf{if} $\mathsf{TEE.Verify}(t,c,\sigma_{\mathrm{BOF}})$ \textbf{then}
        \label{ln:verify-bof-echo}
        $\mathit{placements}[t].\mathsf{add}(c,\sigma_{\mathrm{BOF}})$
            \label{ln:store-bof-echo}
\EndProcedure

\Procedure{AfterCollectionWindow}{$t$} \Comment{$2\Delta$ after first reception of $t$}
    \label{ln:collection-window}
    \State $\mathsf{FinalizeOrderingInfo}(t,\mathit{timestamps}[t])$
        \Comment{OL: local ordering key}
        \label{ln:finalize}
    \State $\mathsf{AddToCurrentCar}(t, K_r(t))$ \Comment{OL only}
        \label{ln:car-ol}
    \State $\mathsf{TEE.RecordAnnouncements}(t,\mathit{placements}[t])$
        \Comment{BOF: advances $(w,M)$}
        \label{ln:record-ann}
\EndProcedure
\end{algorithmic}
\end{algorithm}

    \subsection{Synchronous Consensus}
    \label{sec:consensus-layer}

\pname{}'s consensus layer is based on the Sync~HotStuff~\cite{synchotstuff} synchronous algorithm that requires $n \ge 2f{+}1$ replicas, which both \pname{} variants satisfy.
In its steady state, blocks form a hash chain and the leader broadcasts a block
(l.~\ref{ln:broadcast-propose}) with a certificate of $f{+}1$ votes for its predecessor. A replica that has detected no leader equivocation forwards the proposal to all replicas, broadcasts its vote (l.~\ref{ln:send-vote}), and starts a $2\Delta$ timer. It commits the block and its ancestors if the timer expires while it has neither detected equivocation nor quit the view (l.~\ref{ln:commit-timer}).

Forwarding and broadcasting are what make the $2\Delta$ commit rule safe~\cite{synchotstuff}.
Commit timers do not block progress: a replica may vote at the next height without waiting for the previous block's timer to expire.
\pname{} keeps this structure and its commit rule unchanged, and modifies only the content of blocks and how replicas validate them.

\textbf{Propose.} Besides the Sync~HotStuff certificate for its predecessor block, a block proposal contains the PoAs of a \emph{tip cut} $\langle \mathit{tip}_1, \ldots, \mathit{tip}_n \rangle$, i.e., one certified lane tip per replica, and, in \namebof{}, the delinquency claims introduced below (Alg.~\ref{alg:consensus}, ll.~\ref{ln:tipcut}--\ref{ln:claims}).
Eligibility for execution is decided after commitment, from the committed Cars alone (\S\ref{sec:exec-threshold} and \S\ref{sec:bof-eligibility}).
Committing a tip cut transitively commits every Car in the history of each lane tip, so a constant-size proposal references an arbitrarily large set of transactions and dissemination remains decoupled from consensus.

\textbf{Vote.} Before voting for a block, a replica checks that
(i)~the tip cut is \emph{valid}, i.e., every referenced tip carries a valid PoA and no lane regresses below its last committed position, and \emph{fresh}, i.e., every $\mathit{tip}_j$ is at least as high as the highest certified tip of lane~$j$ that the voter \rev{sent to the leader more than $2\Delta$ before it received the proposal} (Alg.~\ref{alg:consensus}, l.~\ref{ln:check-tipcut}), and
(ii)~in \namebof{}, its own delinquency timer has expired for every delinquency claim that the block carries (l.~\ref{ln:wait-timer}).
A replica that fails a check abstains from voting, permanently for~(i), and until its timers expire for~(ii), if they do. It otherwise follows Sync~HotStuff, forwarding the proposal and broadcasting its vote to all replicas (l.~\ref{ln:send-vote}) rather than replying to the leader alone.

Validity alone lets a faulty leader propose well-formed blocks at full rate while possibly pinning a lane at a stale tip. Freshness is what keeps a leader from starving a lane while remaining unblamable.
\rev{A replica that receives a proposal omitting a certified tip that it holds sends the corresponding PoA to the leader. This forwarding is necessary because a faulty lane owner may reveal a certified tip to some replicas only. \lean{For the same reason, a replica that abstains from voting on a block because one of its tips is stale broadcasts the corresponding PoA to all replicas, once per tip and view. Otherwise a correct replica that is the only one to hold the tip would abstain from every block of a faulty leader that omits it, while the other replicas would keep certifying these blocks, and it would never vote, hence never commit, again in this view. With the broadcast, every correct replica learns the tip, forwards it to the leader, and abstains in turn $2\Delta$ later, so the leader must reference the tip or lose every correct vote. A correct leader never triggers this broadcast.} The freshness check never rejects a correct leader. A PoA that a voter sent more than $2\Delta$ before receiving a proposal reached the leader within~$\Delta$, hence before the leader sent this proposal, which took at most~$\Delta$ to arrive. \lean{A leader cannot omit a certified tip for long. Once a correct replica votes for a block that omits a tip held by all correct replicas, they all receive this block within~$\Delta$ and have sent the PoA to the leader by then. No correct replica votes for a block that still omits the tip if it receives this block more than $3\Delta$ after that vote, so no such block is certified.}}

\lean{\textbf{Relation to Sync~HotStuff.} These checks only make a correct replica abstain from proposals that Sync~HotStuff would let it vote for. They never make it vote for a proposal that Sync~HotStuff rejects, so \pname{} inherits the safety of Sync~HotStuff~\cite{synchotstuff}. It does not inherit its liveness, because a replica that abstained from the proposals of a correct leader would eventually blame it. The checks, however, accept every proposal that a correct leader makes by the rules above, at every correct replica. A correct leader references every certified tip that reaches it before it proposes, and a PoA that a voter sent more than $2\Delta$ before receiving the proposal reached the leader in time. A correct leader's own timers likewise bound those of the voters within~$\Delta$. No correct replica therefore blames a correct leader, the view of a correct leader never ends, and a view that never ends keeps committing blocks, whether its leader is correct or not. Our proofs take the commit rule and the view-change rules of Sync~HotStuff as given.}

\textbf{Lane delinquency.} \namebof{} must also be able to conclude that a lane will not deliver a Car that its enclave has announced (\S\ref{sec:bof-eligibility}).
To do so, a block may carry \emph{delinquency claims} $\langle j, c\rangle$, asserting that lane~$j$ has not reached position~$c$, which check~(ii) of the \emph{Vote} paragraph subjects to the voter's own timer. A claim names the lowest $\hat{c}_j(t)$ (\S\ref{sec:bof-eligibility}) of a pending transaction that lane~$j$'s committed prefix has not reached. A voter also abstains from a block that carries neither this claim nor a tip of lane~$j$ at this position or beyond, if it armed its timer of duration $D = 5\Delta$ for the claim more than $D{+}4\Delta$ before it received the block \lean{(l.~\ref{ln:late-rule})}, so that a leader cannot ignore a delinquent lane. \lean{This obligation ends once a committed claim names lane~$j$. It never rejects a correct leader, which arms its own timer at most~$\Delta$ after the voter and thus proposes such a block more than $2\Delta$ after that timer expired, when it carries the claim or references the Car.} A block certificate carries $f{+}1$ votes, hence a correct one, so a committed claim is always backed by a correct replica that has waited long enough; and because the claim is committed, all correct replicas act on the same claims. A claim that no correct replica supports gathers no vote.
Check~(ii) of the \emph{Vote} paragraph never delays a correct replica's vote on a correct leader's proposal (\S\ref{sec:bof-eligibility}), so \pname{} keeps Sync~HotStuff's view-change procedure unchanged.

\textbf{View changes.} Replicas leave a view when the leader equivocates or stalls. \pname{} follows the view-change rules of Sync~HotStuff unchanged. \lean{A replica broadcasts a blame message for view~$v$ if fewer than~$p$ proposals have triggered its vote within $(2p{+}4)\Delta$ in this view, where $p$ is a parameter of the protocol, or if it detects that the leader equivocates, in which case it attaches the two conflicting proposals. A replica that gathers $f{+}1$ blame messages for view~$v$ broadcasts them, stops voting in view~$v$, and aborts its commit timers of this view. It then waits $2\Delta$, locks on the highest certified block that it knows, sends this block and its certificate to the leader of view~$v{+}1$, and enters view~$v{+}1$. The new leader waits $2\Delta$ after entering the view and proposes a block that extends the highest certified block it has received. A replica forwards this proposal and votes for it if the certified block it extends ranks at least as high as its own lock, after which the steady state resumes.} \lean{A new leader builds its proposals like any other leader. Its tip cut references the certified lane tips that it holds, and in \namebof{} it carries the delinquency claims that its own timers allow. Nothing is inherited from the previous leader. In particular, a PoA that a replica forwarded to the previous leader is forwarded again to the new one, on the first of its proposals that omits the tip.}

\begin{algorithm}[t]
\caption{\pname{} consensus layer (steady state): block proposal and validation}
\label{alg:consensus}
\algfont
\begin{algorithmic}[1]

\Procedure{RunView}{$\mathit{view}$} \Comment{at the leader}
    \State $\mathit{tipcut} \gets
        \langle \mathit{tip}_1,\ldots,\mathit{tip}_n\rangle$
        \Comment{latest certified lane tips}
        \label{ln:tipcut}
    \State $\mathit{claims} \gets \{\langle j,c\rangle :
        \text{lane } j \text{ delinquent at } c\}$
        \Comment{BOF only}
        \label{ln:claims}
    \State $\mathit{block} \gets
        \mathsf{TEE.SignBlock}(\mathit{tipcut},\mathit{claims})$
        \label{ln:sign-block}
    \State $\mathsf{Broadcast}(\textsc{Propose},\mathit{block})$
        \label{ln:broadcast-propose}
\EndProcedure

\Procedure{OnReceivePropose}{$\mathit{block}$} \Comment{at every replica}
    \If{$\neg\mathsf{ValidTipCut}(\mathit{block}.\mathit{tipcut})$}
        \label{ln:check-tipcut}
        \State \lean{$\mathsf{BroadcastStaleTips}(\mathit{block}.\mathit{tipcut})$}
            \Comment{\lean{PoAs, to all replicas}}
            \label{ln:broadcast-stale}
        \State \Return
        \Comment{invalid PoA, lane regression, or stale tip}
    \EndIf
    \State $\mathsf{ForwardOmittedTips}(\mathit{block}.\mathit{tipcut})$
        \Comment{PoAs, to the leader}
        \label{ln:forward-poa}
    \State \lean{\textbf{if} some timer $\langle j,c\rangle$ was armed more than $D{+}4\Delta$ ago, no committed claim names lane~$j$, and $\mathit{block}$ has neither the claim $\langle j,c\rangle$ nor a tip of lane~$j$ at position $\geq c$ \textbf{then return}}
        \Comment{\lean{BOF only}}
        \label{ln:late-rule}
    \State \textbf{wait until} the delinquency timer of every
        claim of $\mathit{block}$ has expired
        \Comment{BOF only}
        \label{ln:wait-timer}
    \State $\mathsf{Broadcast}(\textsc{Propose},\mathit{block})$
        \Comment{Proposal forwarding}
        \label{ln:forward-propose}
    \State $\mathsf{Broadcast}(\textsc{Vote},\mathit{block})$
        \label{ln:send-vote}
    \State $\mathsf{StartCommitTimer}(\mathit{block}, 2\Delta)$
        \label{ln:start-timer}
\EndProcedure

\Procedure{OnCommitTimerExpiry}{$\mathit{block}$}
    \label{ln:commit-timer}
    \If{leader equivocation detected \textbf{or} view left}
        \State \Return
        \Comment{Sync~HotStuff's commit rule}
    \EndIf
    \State $\mathsf{UpdateOrderingState}(\mathit{block}$ and its ancestors$)$
        \label{ln:update-ordering}
    \State $\mathsf{ExecuteEligibleTransactions}()$
        \label{ln:execute}
        \Comment{see \S\ref{sec:impl-ol} or \S\ref{sec:impl-bof}}
\EndProcedure
\end{algorithmic}
\end{algorithm}

\section{\nameol{}: Ordering Linearizability}
\label{sec:impl-ol}

In \nameol{},
the ordering indicator that a TEE attaches to a transaction is its host's clock reading when the transaction reaches the enclave, which the enclave attests and checks to exceed the last timestamp it signed.
Every final ordering key falls within the range of its correct timestamps.

    \subsection{From Timestamps to Ordering Keys}
    \label{sec:ordering-logic}

\textbf{Local ordering key.} Once its $2\Delta$ collection window for a transaction~$t$ has closed, replica~$r$ forwards the timestamps it collected for $t$ to its TEE. The TEE verifies that they are signed for~$t$ by at least $f{+}1$ distinct enclaves, selects the $(f{+}1)$-th smallest timestamp, and certifies the selected value.
We call this value the \emph{local ordering key} $K_r(t)$ of~$t$ at replica~$r$.
While local ordering keys eliminate deflation attacks, where faulty replicas give low timestamps to a transaction, a faulty replica can still inflate its own local ordering key by withholding some ordering indicators from its TEE. \nameol{} therefore computes final ordering keys from several local ordering keys.

\textbf{Final ordering key.}  Once a replica has obtained a local ordering key for a transaction, it inserts the transaction along with the local ordering key in a Car it generates and disseminates, so that it is eventually committed  by the consensus protocol. Since at most~$f$ replicas are faulty, at most~$f$ local ordering keys may be inflated.
The \emph{final ordering key} of a transaction~$t$ is the minimum over the first $f{+}1$ local ordering keys for~$t$ that are committed:
\(
    K(t) = \min_{r \in S_{f+1}(t)} K_r(t),
\)
where $S_{f+1}(t)$ is the set of replicas whose local ordering keys for~$t$ are committed first (decided by block height, then by replica identifier, then by position within the block).
By Lemma~\ref{lem:key-range}, this is the key that Lemma~\ref{lemma:sync-ol} requires.

\begin{lemma}[Key Range]
\label{lem:key-range}
For every transaction~$t$, the final ordering key satisfies
\(
    \min_{c \in C} T_c(t) \le K(t) \le \max_{c \in C} T_c(t),
\)
where $C$ is the set of correct replicas.
\end{lemma}

\begin{proof}
\emph{Lower bound (deflation protection).}
Every local ordering key, including one certified for a faulty replica, is the $(f{+}1)$-th smallest element of a set of genuine TEE attestations: a TEE produces exactly one signed timestamp per transaction (\S\ref{sec:tee-indicators}), so a faulty host cannot forge low timestamps or present different ones to different replicas; it can only withhold attestations, which never decreases the $(f{+}1)$-th smallest of the submitted set.
Among the $f{+}1$ smallest attestations submitted to a TEE, at least one originates from a correct replica, since at most~$f$ replicas are faulty.
The $(f{+}1)$-th smallest submitted timestamp is therefore at least that correct timestamp, so $K_r(t) \ge \min_{c \in C} T_c(t)$ for every replica~$r$, and the minimum over the keys of $S_{f+1}(t)$ preserves this bound.

\emph{Upper bound (inflation protection).}
The set $S_{f+1}(t)$ contains $f{+}1$ distinct replicas, so at least one of them, say~$r^*$, is correct.
A correct replica computes its local ordering key only once its $2\Delta$ collection window has closed (\S\ref{sec:data-layer}), by which time it has collected the attestations of all $f{+}1$ correct replicas.
A set that contains all $f{+}1$ correct timestamps has its $(f{+}1)$-th smallest element at most the largest of them, so $K_{r^*}(t) \le \max_{c \in C} T_c(t)$, and taking the minimum gives $K(t) \le K_{r^*}(t) \le \max_{c \in C} T_c(t)$. \qed
\end{proof}

    \subsection{Execution Threshold and Finality}
    \label{sec:exec-threshold}

Computing $K(t)$ is necessary but not sufficient to execute a transaction. A replica must also ensure that no transaction whose key is not yet computed can be ordered before it.
\nameol{} obtains this guarantee with an \emph{execution threshold}~$\tau$: a lower bound, computable from the committed Cars alone, on the smallest final ordering key that any transaction whose key is not yet fixed might still obtain.

The threshold is read off the completion frontiers of \S\ref{sec:data-layer}. Let $G_j$ be the largest frontier sealed into a committed Car of lane~$j$; like every quantity in this subsection except the issued timestamps themselves, it is a function of the committed prefix of Cars, and it grows as further Cars commit. An enclave never advances its frontier past a transaction whose key its host has withheld, nor past the largest indicator it has issued (\S\ref{sec:data-layer}). Consequently, a transaction that $j$ attested at or below $G_j$ already has its key in a committed Car, and one that $j$ has not yet attested will, by enclave monotonicity, receive an indicator larger than~$G_j$. No transaction can therefore sit below a committed frontier and still be waiting to appear.

For a transaction~$t$ and a replica~$j$, define
\[
    \underline{T}_j(t) = \begin{cases}
        G_j^{(i-1)} & \text{if $j$ \lean{first} published $K_j(t)$ in Car~$i$,} \\[2pt]
        G_j         & \text{otherwise,}
    \end{cases}
\]
where $G_j^{(i-1)}$ is the frontier sealed into Car~$i{-}1$ of lane~$j$\lean{, and Car~$i$ is the first committed Car of lane~$j$ that carries $K_j(t)$. Any later Car that repeated the key could carry a frontier above $T_j(t)$}. Let $\underline{K}(t)$ denote the $(f{+}1)$-th smallest of $\underline{T}_1(t),\ldots,\underline{T}_n(t)$, and let $\mathcal{U}$ be the set of transactions that appear in the committed Cars with fewer than $f{+}1$ local ordering keys. A transaction absent from the committed Cars has $\underline{T}_j(t) = G_j$ for every~$j$, so the single term $G^{(f+1)}$, the $(f{+}1)$-th smallest of $G_1,\ldots,G_n$, accounts for all of those at once. The threshold is
\begin{equation}
    \tau = \min\Big( G^{(f+1)},\;\; \min_{t \in \mathcal{U}} \underline{K}(t) \Big).
    \label{eq:tau}
\end{equation}
A transaction is eligible for execution as soon as its final ordering key has
been fixed by $f{+}1$ committed local ordering keys and $K(t) < \tau$. Transactions whose key reaches~$\tau$ wait for a later block.
\lean{Eligible transactions are executed by increasing final ordering key, and transactions with equal keys by increasing transaction digest.}
\lean{Lemma~\ref{lem:exec-threshold} in fact places every unfixed key strictly above~$\tau$. The strict inequality only concerns transactions whose key is fixed and equals~$\tau$, which wait for a later block.}

\begin{lemma}[Execution Threshold Safety]
\label{lem:exec-threshold}
At any point of the execution, with~$\tau$ evaluated on the prefix of Cars committed at that point, no transaction whose final ordering key is not yet fixed can ever obtain a final ordering key smaller than~$\tau$\lean{, nor can any replica ever certify for it a local ordering key that is not above~$\tau$}.
\end{lemma}

\begin{proof}

  Let $t'$ be a transaction whose final ordering key is not yet fixed, and let $P(t')$ be the $(f{+}1)$-th smallest of all timestamps ever issued for~$t'$, including those issued after the evaluation point. At least $f{+}1$ correct timestamps eventually exist.
Every local ordering key for~$t'$, whenever it is certified, is the $(f{+}1)$-th smallest of a subset of these timestamps, so the fact above gives $K_r(t') \geq P(t')$ for every replica~$r$, hence $K(t') \geq P(t')$.

  Consider a replica~$j$ that ever issues a timestamp $T_j(t')$.
  If $j$ published $K_j(t')$ in Car~$i$ of the committed prefix, then $t'$'s key was absent from Car~$i{-}1$, so by the frontier's definition $T_j(t') > G_j^{(i-1)} = \underline{T}_j(t')$.
  If $j$ attested~$t'$ but published no key for it in any committed Car, the same reasoning against the latest committed Car of its lane gives $T_j(t') > G_j = \underline{T}_j(t')$.
  If $j$ has not attested~$t'$ by the evaluation point, the indicator it later issues exceeds every indicator it issued before, and its frontier never exceeds its largest issued indicator (\S\ref{sec:data-layer}), so again $T_j(t') > G_j = \underline{T}_j(t')$.
  Every issued timestamp thus exceeds its replica's entry of $\underline{T}(t')$, and replicas that never issue a timestamp for~$t'$ only remove elements from the comparison, so the fact above gives $P(t') \geq \underline{K}(t')$.

  It remains to bound $\underline{K}(t')$. If $t'$ appears in the committed Cars, it lies in~$\mathcal{U}$, because its key is unfixed, and $\underline{K}(t') \geq \tau$ by Eq.~\eqref{eq:tau}. If it does not, then no replica has published a key for~$t'$, so $\underline{T}_j(t') = G_j$ at every~$j$ and $\underline{K}(t') = G^{(f+1)} \geq \tau$.
  In both cases $K(t') \geq P(t') \lean{{}>{}} \underline{K}(t') \geq \tau$\lean{, and the same holds for every local ordering key $K_r(t')$}.
  \qed{}
\end{proof}

    \subsection{Correctness}
    \label{sec:security}

\begin{theorem}
\label{thm:ol-impl}
Under synchrony, \nameol{} ensures ordering linearizability (Definition~\ref{def:ol}) with $2f{+}1$ TEE-equipped replicas.
\end{theorem}

\begin{proof}
\emph{Ordering Linearizability.}
We first argue that the order in which \nameol{} executes transactions is the order of increasing final ordering keys. A transaction is executed only once its key has been fixed by $f{+}1$ committed local ordering keys and lies below the execution threshold, and by Lemma~\ref{lem:exec-threshold} no transaction whose key is still unfixed can ever obtain a key below that threshold. The realized execution order therefore coincides with the order by increasing~$K(\cdot)$. By Lemma~\ref{lem:key-range}, every final ordering key falls within the range of its transaction's correct timestamps, so Lemma~\ref{lemma:sync-ol} applies and this order implements ordering linearizability.
\lean{It remains to show that $t_1$ is not left behind when $t_2$ is executed, for $t_1$ and $t_2$ as in Definition~\ref{def:ol}. Some correct replica eventually closes its collection window for~$t_1$ and certifies a local ordering key at most $\max_{c \in C} T_c(t_1)$ (\S\ref{sec:data-layer}), which is below $K(t_2)$ and hence below the threshold~$\tau$ under which $t_2$ is executed. By Lemma~\ref{lem:exec-threshold}, the key of~$t_1$ is then already fixed at that point. Since $K(t_1) < K(t_2) < \tau$, $t_1$ is executed in the same step or earlier, before~$t_2$.}

\emph{Finality.} The same argument makes execution irrevocable: once the transactions with $K(t) < \tau$ are executed, Lemma~\ref{lem:exec-threshold} guarantees that no later transaction can acquire a final ordering key that requires insertion before this executed prefix.

\emph{Agreement.} Correct replicas commit the same blocks by the safety of Sync~HotStuff, and apply the same deterministic rules.

\emph{Liveness.} Every correct client's transaction is eventually disseminated, its collection window closes, and its local ordering key is committed. $K(t)$ is fixed when $f{+}1$ local ordering keys are committed. The same holds for a transaction that only a faulty replica disseminated. Its Car was acknowledged by a correct replica, which first obtained its body and echoed it (\S\ref{sec:data-layer}), so all correct replicas attest it, close their collection window and publish a local ordering key. The correct replicas keep sealing Cars\lean{, every correct replica holds the PoA of each of these Cars within $4\Delta$ of its seal, and the freshness rule (\S\ref{sec:consensus-layer}) then keeps any leader from omitting it, so every Car of a correct lane is eventually committed}. Since each correct enclave keeps issuing indicators, for client transactions or, when traffic pauses, for the heartbeat transactions of \S\ref{sec:data-layer}, the completion frontiers $G_j$ of the correct replicas grow without bound. \lean{Each correct lane~$j$ thus has a Car~$i_j$ from which its frontiers exceed $K(t)$, because frontiers never decrease (\S\ref{sec:data-layer}). Eventually $G_j > K(t)$ at the $n{-}f$ correct lanes, so $G^{(f+1)} > K(t)$. Under load, however, new transactions keep entering~$\mathcal{U}$, so $\mathcal{U}$ never becomes empty and the second term of Eq.~\eqref{eq:tau} must also exceed $K(t)$. Consider $t' \in \mathcal{U}$ and a correct lane~$j$. If $j$ has not published $K_j(t')$, then $\underline{T}_j(t') = G_j > K(t)$. If it first published it in a Car after~$i_j$, then $\underline{T}_j(t')$ is the frontier of a Car at or after~$i_j$ and also exceeds $K(t)$. Only the finitely many transactions whose key a correct lane published up to Car~$i_j$ remain, and each of them leaves~$\mathcal{U}$ for good once its key is fixed. From then on $\underline{T}_j(t') > K(t)$ at every correct lane, hence $\underline{K}(t') > K(t)$ for every $t' \in \mathcal{U}$,} so $K(t) < \tau$ and $t$ becomes eligible for execution.
\qed{}
\end{proof}

\section{\namebof{}: $\gamma$-Batch-Order Fairness}
\label{sec:impl-bof}

In \namebof{},
the ordering indicator is the value of a counter that the enclave increments at every attestation, so the sequence numbers $s_r(\cdot)$ that replica~$r$'s TEE issues reveal the order in which $r$ received transactions.
Lemma~\ref{sync-bof} shows that with $n > \frac{2\gamma f}{2\gamma - 1}$, $t_a$ can be ordered no later than $t_b$ whenever at least $\gamma(n{-}f)$ of the counted reports place $t_a$ first, over a set of reports that includes those of every correct replica.

    \subsection{Attestations, Settlement, and Eligibility}
    \label{sec:bof-eligibility}

In \namebof{}, transactions appear in Cars in the order they have been observed by a TEE. It is therefore unnecessary to include ordering indicators in Cars. Instead, replicas obtain a single signature on a Car from their TEE, which verifies that the transactions appear in it according to the order it observed them.
A Car committed at position~$i$ of lane~$r$ commits every earlier Car of that lane, so the committed Cars of lane~$r$ hold every transaction $r$ sequenced at or before~$t$, and no transaction can later appear and claim a position before~$t$ in $r$'s local order.

\textbf{Settlement.} Let $c_r(t)$ be the position of the Car of lane~$r$ that contains~$t$ (\S\ref{sec:data-layer}). Lane~$r$ \emph{covers}~$t$ once the committed prefix of that lane contains~$t$ and one of its committed Cars carries a reception frontier with $w_r \geq c_r(t)$. Let $M_r^t$ be the announcement vector of the first such Car. Let $\mathcal{C}_{f+1}(t)$ be the first $f{+}1$ lanes that cover~$t$, ordered by block height, then replica identifier, then position within the block, and let
\(
    \hat{c}_j(t) \;=\; \max_{r \in \mathcal{C}_{f+1}(t)} M_r^t[j] .
\)
At least one lane of $\mathcal{C}_{f+1}(t)$ is correct, and a correct~$r$ seals $M_r[j] \geq c_j(t)$ for every correct~$j$ (\S\ref{sec:data-layer}), so $\hat{c}_j(t) \geq c_j(t)$ at every correct replica~$j$: a faulty lane may inflate the maximum, but no coalition of them can pull it below what a correct lane has sealed.

Lane~$j$ is \emph{settled} for~$t$ once the committed prefix of lane $j$ has reached position $\hat{c}_j(t)$, or once any committed delinquency claim names lane~$j$. In the first case, either $t$ appears in the committed Cars of lane~$j$, which fixes $s_j(t)$, or it does not, and $j$ is faulty. In the second case, a committed delinquency claim implies that a correct replica voted on it, and thus $j$ is faulty. A lane named by a committed claim contributes no ordering report thereafter. \add{The other lanes are \emph{counted}.}
A transaction is \emph{eligible} once every lane is settled for it.

\rev{\textbf{Release across execution steps.} Eligibility is decided per transaction, so two transactions bound by an edge may become eligible in different blocks. For instance, if every correct replica receives $t_a$ before~$t_b$ and the $f$ faulty lanes include $t_b$ but not~$t_a$, then $t_b$ is covered by $f{+}1$ lanes and may be eligible and ordered before $t_a$ is even covered. \namebof{} therefore orders an eligible transaction only once it is \emph{released}, i.e., once the transactions it must wait for are eligible too.}
\add{Let $\theta = \lceil\gamma(n{-}f)\rceil$ be the edge threshold of \S\ref{sec:bof-dep-graph}. A transaction~$v$ \emph{blocks} an eligible~$u$ if $v$ precedes~$u$ in at least $\theta$ counted lanes. Waiting for all blockers, and for their blockers in turn, would not be live, as such chains grow without bound under continuous load. The chain is thus cut where $t$ is \emph{ahead of}~$v$\lean{, for $v \neq t$}, i.e., where at least $\theta$ counted lanes have committed~$t$ with $v$ not before it, which can be decided even before $v$ is committed. The \emph{closure} $D(t)$ of an eligible~$t$, the set that $t$ waits for, is the smallest set that contains~$t$ and, for every eligible $u \in D(t)$, each unexecuted blocker~$v$ of~$u$ that $t$ is not ahead of. Once every transaction of $D(t)$ is eligible, the unexecuted ones are released, and the step orders the released transactions. A blocker left out of $D(t)$ is ordered later, but lies on a cycle with the transaction it blocks (Lemma~\ref{lem:cycle-closing}), i.e., in the same batch.}
\rev{Like eligibility, release is a function of the committed prefix alone, so all correct replicas agree on it.}

\textbf{Suspicion and the delinquency timer.} Let $c = \hat{c}_j(t)$ and
\(
    D = 2X + 3\Delta = 5\Delta.
\)
A replica that has committed the block that fixes~$c$ and has not \emph{received} a certified Car of lane~$j$ at position~$c$ or beyond within~$D$ broadcasts a \textsc{Suspect} message for $\langle j,c \rangle$. This message carries the announcement, signed by $j$'s enclave, that sets~$c$, which proves that this enclave has opened Car~$c$. A replica that receives a valid \textsc{Suspect} message for the first time relays it and arms its delinquency timer for $\langle j,c \rangle$, which also lasts~$D$. Correct replicas thus arm their timers within~$\Delta$ of one another, whenever they commit. A replica that already holds a certified Car of lane~$j$ at position~$c$ or beyond, or that receives one before its timer expires, cancels its timer and broadcasts the corresponding PoA. \lean{A Car received at the very instant the timer expires counts as received in time, since the budget below has no slack.} Otherwise its timer expires and it votes for the claim (check~(ii) of the \emph{Vote} paragraph of \S\ref{sec:consensus-layer}).
A leader proposes $\langle j,c \rangle$ once $2\Delta$ have elapsed since its own timer expired, if it still holds no such Car. It references the Car in its tip cut otherwise. When a correct leader proposes a claim, the timers of all correct replicas have expired and none was cancelled, since the PoA would have reached the leader within these $2\Delta$. Correct replicas therefore vote for it without waiting.
The budget follows the path a correct lane still has to travel: a correct~$j$ has already opened Car $\hat{c}_j(t)$ when the timer is armed, seals it within~$X$, collects its $f{+}1$ acknowledgments $2\Delta$ later, carries the resulting PoA in its next seal at most~$X$ after that, and delivers it one~$\Delta$ afterwards.
No correct replica's timer can therefore expire on a correct lane, whatever the leaders do, so a claim against a correct lane never gathers the correct vote that its certificate requires.

    \subsection{Dependency Graph Construction}
    \label{sec:bof-dep-graph}

Replicas build a dependency graph $G = (V, E)$ whose vertices are the \rev{released} transactions and whose edges encode pairwise ordering decisions.
For each pair $t_a, t_b \in \mathcal{T}$ with $t_a \neq t_b$, let $\mathcal{R}$ be the set of counted lanes (\S\ref{sec:bof-eligibility}) whose committed Cars contain both transactions when the pair is released. A faulty lane may commit a transaction after it was settled for it. Whether $\mathcal{R}$ includes such a lane is immaterial, since Lemma~\ref{lem:edge-correct} holds for any $\mathcal{R}$ that contains the correct lanes.
Both transactions being eligible, every lane is settled for both, so $\mathcal{R}$ contains every correct replica (\S\ref{sec:bof-eligibility}) and $|\mathcal{R}| \geq n{-}f$.
Let $k_{a \locprec b} = |\{R \in \mathcal{R} : s_R(t_a) < s_R(t_b)\}|$ be the number of replicas of~$\mathcal{R}$ that received $t_a$ first,
and let $k_{b \locprec a} = |\mathcal{R}| - k_{a \locprec b}$ be the complementary count.
With the edge threshold $\theta = \lceil \gamma (n{-}f) \rceil$, which mirrors the quorum of Lemma~\ref{sync-bof}, the edge rule is
\begin{equation}
\label{eq:edge-rule}
    \text{if } k_{a \locprec b} \geq \theta \text{, add } t_a \to t_b
    \text{; if } k_{b \locprec a} \geq \theta \text{, add } t_b \to t_a.
\end{equation}
If neither count reaches~$\theta$, no edge is added and the relative placement of the two transactions is left to the topological order of \S\ref{sec:bof-ordering} (i.e., they might not be batched together).
\rev{By Lemma~\ref{lem:edge-correct} this happens only for pairs on which fewer than $\gamma(n{-}f)$ correct replicas agree either way.}
\rev{Note that, when $\gamma = 1$ and $n = 2f{+}1$, the rule requires $f{+}1$ concurring attestations out of the up to $2f{+}1$ that $\mathcal{R}$ may contain.}

\begin{lemma}[Edge direction correctness]
\label{lem:edge-correct}
In a synchronous network with $n > \frac{2\gamma f}{2\gamma - 1}$ replicas,
at most one of the two edges of rule~(\ref{eq:edge-rule}) is added for a given pair. If at least $\gamma(n{-}f)$ correct replicas receive~$t_a$ before~$t_b$, then that edge is $t_a \to t_b$\rev{, and every edge $t_a \to t_b$ that the rule adds is supported by at least $\theta {-} f$ correct replicas that received $t_a$ before~$t_b$}.
\end{lemma}

\begin{proof}
\emph{No wrong edge.} The replicas of $\mathcal{R}$ reporting $t_b \locprec t_a$ are at most the $(1{-}\gamma)(n{-}f)$ correct replicas that received $t_b$ first and the~$f$ faulty ones.
The proof of Lemma~\ref{sync-bof} shows that $n > \frac{2\gamma f}{2\gamma - 1}$ implies $(1 - \gamma)(n - f) + f < \gamma(n - f) \leq \theta$,
so $k_{b \locprec a} < \theta$ and the edge $t_b \to t_a$ is not added. This holds whatever attestations $\mathcal{R}$ happens to contain. Exclusivity follows from counting instead, and for every pair: $k_{a \locprec b} + k_{b \locprec a} = |\mathcal{R}| \leq n$, whereas $2\theta \geq 2\gamma(n{-}f) > n$ by the same replica bound, so the two counts cannot both reach~$\theta$.

\emph{The mandated edge.} By \S\ref{sec:bof-eligibility}, $\mathcal{R}$ contains every correct replica, so the $\gamma(n{-}f)$ correct replicas that received $t_a$ first all contribute to $k_{a \locprec b}$, giving $k_{a \locprec b} \geq \theta$
and hence the edge $t_a \to t_b$.
Faulty hosts cannot contradict these reports: they may manipulate when they forward a transaction to their TEE, but non-equivocation makes every committed attestation a genuine and unique statement of its enclave's receive order.

\rev{\emph{Edge support.} The rule adds $t_a \to t_b$ only when $k_{a \locprec b} \geq \theta$. At most~$f$ replicas of $\mathcal{R}$ are faulty, so at least $\theta - f = \lceil \gamma(n{-}f) \rceil - f$ of the counted reports come from correct replicas that received $t_a$ before~$t_b$. Since $\theta > f$, no coalition of faulty replicas can create an edge on its own.} \qed{}
\end{proof}

\add{We show that skipping the blockers $t$ is ahead of preserves fairness, and that $D(t)$ is finite, which liveness requires.}

\add{\begin{lemma}[Cycle closing]
\label{lem:cycle-closing}
Let $t$ be eligible, $u \in D(t)$, and $v$ be a blocker of~$u$\lean{, distinct from~$t$,} that $t$ is ahead of. Then $t$, $v$ and~$u$ lie on a directed cycle in which each edge is supported by at least $\theta {-} f$ correct replicas. Moreover, $t$ is ahead of every transaction that correct enclaves first attest more than~$\Delta$ after they first attest~$t$.
\end{lemma}}

\add{\begin{proof}
By construction of $D(t)$, a chain of blockers leads from~$u$ to~$t$. With $t$ ahead of~$v$ and $v$ blocking~$u$, it closes the cycle $t \to v \to u \to \cdots \to t$. Each of these relations holds in $\theta$ counted lanes, of which at most~$f$ are faulty, and a correct lane reflects the order in which its replica received the two transactions.
For the second claim, every correct enclave attests~$t$ within~$\Delta$ of the first one (\S\ref{sec:data-layer}), hence before such a transaction. Since $t$ is eligible, the $n{-}f \geq \theta$ correct lanes have committed~$t$, and all of them are counted. \qed{}
\end{proof}}

    \subsection{Fair Ordering via SCC Decomposition}
    \label{sec:bof-ordering}

Even correct replicas may observe transactions in different orders, which creates Condorcet cycles in~$G$. We resolve them as Themis~\cite{kelkar2023themis} does (Alg.~\ref{alg:bof-ordering}).
The strongly connected components (SCCs) of~$G$, computed with Tarjan's algorithm~\cite{tarjan1972depth}, form the batches.
The condensation DAG~$G^*$, which has one vertex per SCC and an edge $S_i \to S_j$ whenever $G$ has an edge from a transaction of~$S_i$ to one of~$S_j$, is then topologically sorted to obtain the sequence of batches.
\rev{Every replica derives the batches independently from the committed prefix, so} ties in the sort and the order of transactions within a batch are broken deterministically\lean{, by a key that the submitter of a transaction cannot predict}.

\lean{\textbf{Unpredictable tie-break.} A client can choose the digest of its transaction by varying an unused field, so a digest order would let it place its transaction first in any batch it shares. Ties are therefore broken with a seed derived inside the enclaves. At setup, the enclaves authenticate each other by remote attestation and agree on a key~$s$ that never leaves protected memory. When a committed block~$B$ triggers an execution step, an enclave computes the seed $\rho_B = \mathsf{PRF}_s(H(B))$, but only once its host presents a certificate of $f{+}1$ votes for~$B$, which the enclave checks against the replicas' public keys. The transactions released by~$B$ are then ordered by increasing $H(\rho_B \,\|\, \mathsf{id}(t))$, and a batch is ranked among those it is incomparable with by the smallest such value among its transactions. All enclaves hold the same key and all correct replicas commit the same block, so every replica obtains the same order without exchanging any message. A replica may also take~$\rho_B$ from a peer, together with that enclave's signature on $(H(B), \rho_B)$. Since a certificate requires the vote of a correct replica, a faulty host cannot use its enclave to evaluate the seed of candidate blocks, and every transaction released by~$B$ was submitted before any party could learn~$\rho_B$. A faulty leader obtains a certificate for a second block only by equivocating, which correct replicas detect. If an enclave is compromised and reveals~$s$, the order remains deterministic and agreed upon, and only this tie-break becomes predictable.}

\begin{algorithm}[t]
\caption{\namebof{}: dependency graph and fair ordering}
\label{alg:bof-ordering}
\algfont
\begin{algorithmic}[1]
\Require \rev{Released} transactions $\mathcal{T}$, committed per-replica
    attestations, fairness parameter $\gamma$
\Ensure Ordered sequence of batches $\mathcal{O}$
\State $V \gets \mathcal{T}$;\quad $E \gets \emptyset$;\quad
    $\theta \gets \lceil \gamma \cdot (n{-}f) \rceil$
    \Comment{Edge threshold}
\For{each pair $(t_a, t_b) \in \mathcal{T} \times \mathcal{T}$
    with $t_a \neq t_b$}
    \State $\mathcal{R} \gets$ counted lanes that committed $t_a$ and $t_b$
    \State $k_{a \locprec b} \gets
        |\{R \in \mathcal{R} :
        s_R(t_a) < s_R(t_b)\}|$
    \If{$k_{a \locprec b} \geq \theta$}
        \State $E \gets E \cup \{t_a \to t_b\}$
    \ElsIf{$|\mathcal{R}| - k_{a \locprec b} \geq \theta$}
        \State $E \gets E \cup \{t_b \to t_a\}$
    \EndIf
\EndFor
\State $\{S_1, \ldots, S_k\} \gets \mathsf{TarjanSCC}(V, E)$
    \Comment{Condorcet cycles}
\State $G^* \gets$ condensation DAG of $(V,E)$ over
    $\{S_1, \ldots, S_k\}$
\State $\langle S_{\pi(1)}, \ldots, S_{\pi(k)} \rangle \gets
    \mathsf{DeterministicTopoSort}(G^*)$ \Comment{\lean{ties by seed $\rho_B$}}
\State \Return $\mathcal{O} \gets
    [\mathsf{SortBySeed}(S_{\pi(1)}), \ldots,
    \mathsf{SortBySeed}(S_{\pi(k)})]$
\end{algorithmic}
\end{algorithm}

\begin{figure*}[t]
\centering
\includegraphics[width=\textwidth]{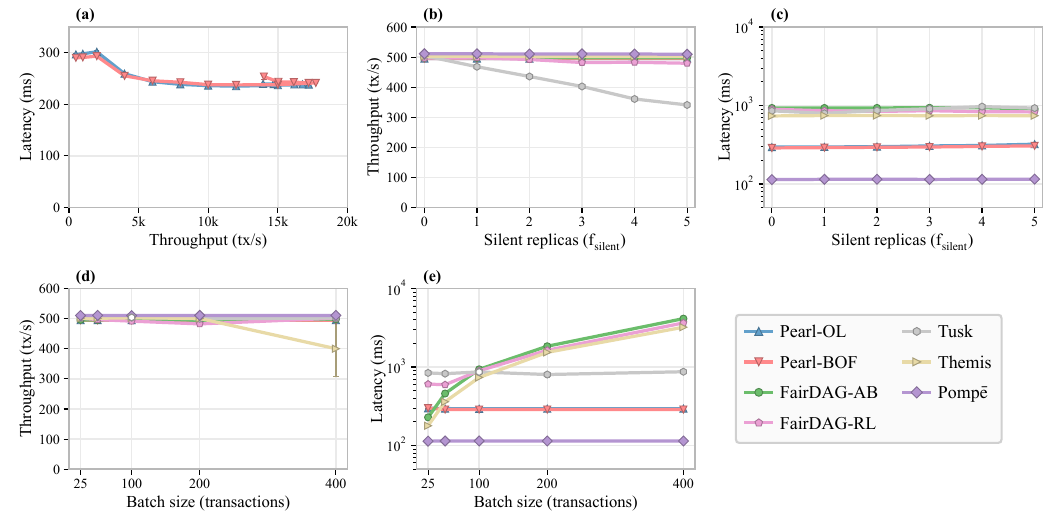}
\caption{(a)~\pname{}'s latency against throughput at the largest committee we evaluate, $n=15$ ($f=7$).
(b)--(c)~Robustness to silent replicas ($f=5$, $500$~tx/s input): throughput and latency as the number of silent replicas varies from $0$ to $5$.
(d)--(e)~Batch-size sensitivity ($f=5$, $500$~tx/s input), without faulty replicas.
\rev{The latency axes of (c) and (e) are logarithmic.}}
\Description{Five line charts on two rows. The first reports latency against achieved throughput for Pearl's OL and BOF modes on a committee of 15 replicas. The next two compare seven systems as the number of silent replicas grows, reporting throughput and latency. The last two do the same as the batch size grows.}
\label{fig:pearl-eval-strip}
\end{figure*}

    \subsection{Correctness}
    \label{sec:bof-correctness}

\begin{theorem}
\label{thm:bof-impl}
In a synchronous network of $n > \frac{2\gamma f}{2\gamma - 1}$ TEE-equipped replicas,
\namebof{} guarantees $\gamma$-batch-order fairness (Definition~\ref{def:bof}).
\end{theorem}

\begin{proof}
Consider transactions $t_a$ and $t_b$ such that at least $\gamma(n{-}f)$ correct replicas receive $t_a$ before~$t_b$.
By Lemma~\ref{lem:edge-correct}, $G$ contains edge $t_a \to t_b$ and not its reverse.
If $t_a$ and $t_b$ belong to two distinct SCCs $S_i$ and $S_j$, then $G^*$ contains the edge $S_i \to S_j$, so the topological sort places $S_i$ before~$S_j$ and $t_a$ is ordered before~$t_b$.
If they belong to the same SCC, they are placed in the same batch,
which satisfies Definition~\ref{def:bof} since it only requires $t_a$ to be ordered no later than~$t_b$.

\rev{\emph{Across execution steps.} The argument above applies to two transactions released in the same step. Suppose $t_b$ is released in some step. Being eligible, $t_b$ has been committed by every correct lane, and the at least $\gamma(n{-}f)$, hence at least~$\theta$, correct replicas that received $t_a$ first hold $t_a$ at an earlier position of their committed lane. No correct lane is named by a committed claim, so $t_a$ } \add{blocks~$t_b$. Let $D(t)$ be a closure with which $t_b$ is released. If $t_a$ has already been executed, it is ordered before~$t_b$. If it is released in the same step, the argument above applies. Otherwise $t$ is ahead of~$t_a$\lean{, with $t_a \neq t$ since $t \in D(t)$ is released in this step}, and by Lemma~\ref{lem:cycle-closing} $t_a$ and~$t_b$ lie on a cycle whose edges are each supported by $\theta{-}f$ correct replicas, i.e., in the same batch.}

\rev{\emph{Batch minimality.} Alg.~\ref{alg:bof-ordering} places two transactions in the same batch only when they lie in the same SCC of~$G$.
By Lemma~\ref{lem:edge-correct}, every edge of the cycle that carries them is supported by at least $\theta - f$ correct replicas\add{, as is every edge of the cycles of Lemma~\ref{lem:cycle-closing}, which join batches across steps}. \namebof{} therefore satisfies batch minimality at support $q = \lceil \gamma(n{-}f) \rceil - f$, which is $n{-}2f$ when $\gamma = 1$.}
Correct replicas commit the same blocks by the safety of Sync~HotStuff, hence observe the same reception frontiers and the same delinquency claims, derive the same eligible \rev{and released sets} (\S\ref{sec:bof-eligibility})
and the same attestations.
TEE non-equivocation makes these attestations unique, so the deterministic rules of Alg.~\ref{alg:bof-ordering} yield the same batch sequence.

\emph{Liveness.} \lean{Every transaction is eventually covered by $f{+}1$ lanes. Every Car of a correct lane is eventually committed, as in \nameol{}, and a correct lane seals a frontier $w_r \geq c_r(t)$ once its collection window for~$t$ has closed, so the $n{-}f \geq f{+}1$ correct lanes cover~$t$.} Every lane is \lean{then} eventually settled for every transaction. \lean{The argument does not rely on a change of leader, since a faulty leader that keeps committing blocks is never replaced. Let $c = \hat{c}_j(t)$ and distinguish two cases. If some correct replica arms its timer for $\langle j, c\rangle$, it relays the \textsc{Suspect} message and every correct replica arms its own within~$\Delta$. Every block that a correct replica receives more than $D{+}4\Delta$ after arming carries the claim or a tip of lane~$j$ at~$c$ or beyond, or gathers no correct vote. Blocks keep being committed, so the claim is committed or the committed prefix reaches~$c$. Otherwise no correct replica ever arms its timer. Each of them then holds a certified Car at~$c$ or beyond within~$D$ of committing the block that fixes~$c$, and the freshness rule (\S\ref{sec:consensus-layer}) keeps any leader from omitting it, so the committed prefix reaches~$c$. For a correct lane no claim is ever committed, since no correct timer expires on it (\S\ref{sec:bof-eligibility}), so its committed prefix reaches~$c$ in both cases. Both cases are bounded in time rather than in views. In the first, every block that a correct replica receives more than $D{+}5\Delta$ after the first correct replica armed its timer carries the claim or the tip. In the second, once a correct replica has voted for a block that omits the Car, no correct replica votes for a block that still omits it and that it receives more than $3\Delta$ after that vote. Lane~$j$ is therefore settled by the first block committed after these delays.} Thus, the eligible set grows\rev{. } \add{Let $t$ be eligible. Every other transaction of $D(t)$ blocks one of its members, hence precedes it in $\theta > f$ lanes, one of them correct, and $t$ is not ahead of it. By Lemma~\ref{lem:cycle-closing}, correct enclaves first attest it at most~$\Delta$ after~$t$. Correct enclaves attest finitely many transactions in a bounded interval, so $D(t)$ is finite. \lean{More precisely, $D(t)$ may change from step to step, but it stays within the finite set of transactions that a correct enclave first attests at most~$\Delta$ after~$t$.} Each of \lean{these} transactions is committed and, by the same argument, eventually eligible\lean{, and eligibility is never lost. From the step at which they are all eligible,} $t$ is released}\rev{, so every committed transaction is eventually ordered.}

\qed{}
\end{proof}

\section{Performance Evaluation}
\label{sec:eval}

\textbf{Artifact.} Our C++ implementations of \nameol{} and \namebof{}, built on ResilientDB~\cite{gupta2020resilientdb}, are public together with raw results and plotting scripts.\footnote{\url{https://github.com/VianRobotin/incubator-resilientdb/tree/vian}}
\rev{We use Intel SGX enclaves~\cite{costan2016sgxexplained} in simulation mode. Enclaves authenticate and verify TEE evidence with AES-128-CMAC under a shared committee key, which remote attestation would provision.}

\textbf{Testbed and workload.} We use the DAS-5 cluster~\cite{bal2016medium}, whose machines share a $1$~Gbps LAN with sub-millisecond RTT.
Each node hosts one replica and one open-loop client that broadcasts $128$B transactions to all replicas. \rev{Tusk's client instead sends each transaction to one replica.}
\rev{We set $\Delta = 50$~ms}, use a block size of $100$ by default, and vary input rates from $500$ to $50{,}000$~tx/s.
\rev{Our prototype applies admission control: each replica sheds the client transactions beyond the rate that keeps its link at $75\%$ of its capacity, so that overload does not push message delays past~$\Delta$. A shed transaction must be resubmitted.}
\rev{Each replica measures the latency of a transaction from its first arrival from a client to its execution, between $15$ and $60$~s after the start of a run, and we report the median over the replicas. Each configuration is run 5 times.}

\textbf{Systems.} We vary $f\in\{3,5,7\}$ and set $\gamma=1$. \pname{} uses $n=2f{+}1$ replicas.
Pomp\=e~\cite{zhang2020byzantine} and FairDAG-AB~\cite{kang2026fairdagconsensusfairnessmultiproposer} (ordering linearizability), FairDAG-RL ($\gamma$-batch-order fairness) and Tusk~\cite{danezis2022narwhal} (no fairness guarantee) use $3f{+}1$ replicas, and Themis~\cite{kelkar2023themis} ($\gamma$-batch-order fairness) uses $4f{+}1$.

\textbf{Scaling and Saturation.} \rev{Fig.~\ref{fig:pearl-eval-strip}(a) plots latency against throughput at $n=15$.
Both modes execute the offered load up to a peak of $40$k~tx/s at $n=7$, $25$k (OL) and $26$k~tx/s (BOF) at $n=11$, and $17$k and $18$k~tx/s at $n=15$.
\namebof{} matches \nameol{} because its graph construction runs off the consensus path.
Latency stays between $226$ and $302$~ms at every load and committee size, close to the $4\Delta = 200$~ms imposed by the collection window and the commit timer.}

\begin{figure*}[!htbp]
\centering
\includegraphics[width=0.98\textwidth]{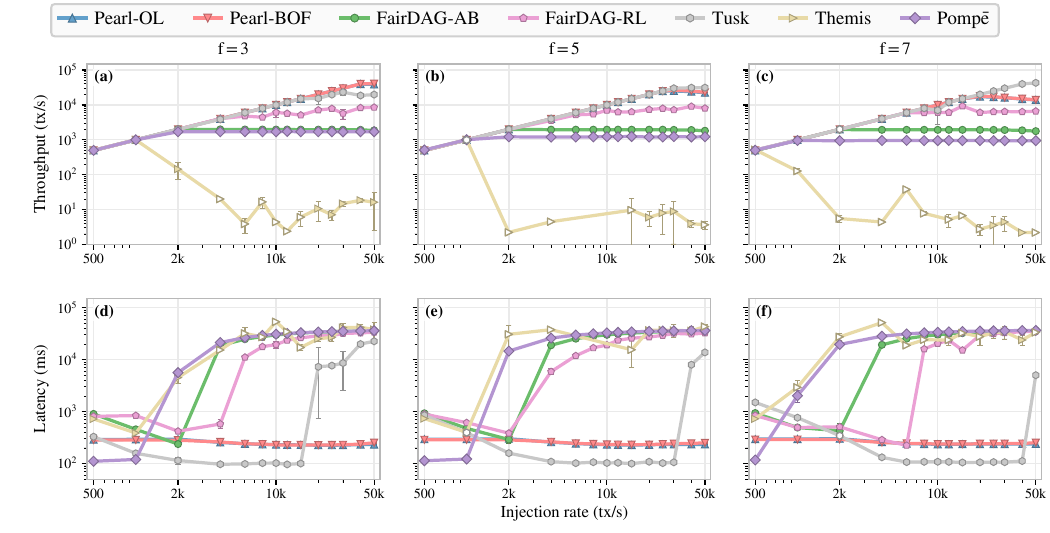}
\caption{Throughput and latency with $f\in\{3,5,7\}$ (columns).
Panels (a)--(c) show throughput versus injection rate and panels (d)--(f) show latency versus injection rate.
All axes are logarithmic.
}
\Description{Six logarithmic line charts compare Pearl-OL, Pearl-BOF, FairDAG-AB, FairDAG-RL, Tusk, Themis,
and Pompe at fault tolerances 3, 5, and 7. The rows show throughput versus load and latency versus load.}
\label{fig:pearl-baseline-overview}
\end{figure*}

\textbf{Comparison with Baselines.} \rev{Fig.~\ref{fig:pearl-baseline-overview} compares \pname{} with the five baselines. We take as the capacity of a system the highest throughput it sustains below $1$~s of latency.
\pname{} reaches $40$k, $26$k and $18$k~tx/s at $f=3$, $5$ and $7$.
FairDAG-AB reaches $2.0$k~tx/s at every~$f$, Pomp\=e $1.7$k, $1.2$k and $0.95$k~tx/s, FairDAG-RL $4.0$k, $2.0$k and $6.0$k~tx/s, and Themis $1{,}000$~tx/s at $f\le5$ and $500$~tx/s at $f=7$.
\pname{} thus sustains $10\times$, $13\times$ and $3\times$ the throughput of the strongest fair-ordering baseline.
At low load, Pomp\=e has the lowest latency ($120$~ms), below \pname{}'s $4\Delta$.}
\rev{Tusk, which provides no fair ordering, sustains $18$k, $30$k and $40$k~tx/s at $100$~ms.
\pname{} reaches twice its throughput at $f=3$, and $84\%$ and $41\%$ of it at $f=5$ and $7$.}

\textbf{Robustness to Silent Replicas.} Each system runs at the smallest $n$ that tolerates $f\!=\!5$ faults ($n\!=\!11$ for \pname{}, $16$ for Pomp\=e, FairDAG and Tusk, $21$ for Themis) at $500$~tx/s, with up to five silent replicas (Fig.~\ref{fig:pearl-eval-strip}(b)--(c)).
\rev{
\pname{}'s latency rises by $7$--$9\%$, from $294$ to $320$~ms (OL) and from $288$ to $307$~ms (BOF), because a silent replica only withholds evidence that the $2\Delta$ collection window already waits for.
The baselines' latency is flat, between $115$~ms (Pomp\=e) and $940$~ms (FairDAG-AB).
Tusk loses the transactions sent to silent replicas, and its throughput falls from $470$ to $340$~tx/s.}

\textbf{Sensitivity to Batch Size.} We sweep the batch size over $\{25, 50, 100, 200, 400\}$ without faults (Fig.~\ref{fig:pearl-eval-strip}(d)--(e)).
\rev{\pname{}'s latency is flat at $300$~ms, because a replica seals a Car once it is full or after~$\Delta$.
FairDAG-AB, FairDAG-RL and Themis wait for a full block, so their latency grows from $230$~ms to $4.1$~s, from $0.6$ to $3.6$~s, and from $180$~ms to $3.2$~s.
Tusk and Pomp\=e are flat, at $0.85$~s and $115$~ms.}

\section{Related Work}
\label{sec:related-work}

\textbf{Fair transaction ordering.}
Pomp\=e~\cite{zhang2020byzantine} introduced ordering linearizability, separating indicator aggregation from consensus and requiring $3f{+}1$ replicas under partial synchrony. AOAB~\cite{gramoli2024aoab} realises the same property with two additional communication rounds.
\pname{} instead binds indicators to their issuing replicas through TEE attestations and exploits synchrony, achieving ordering linearizability with $2f{+}1$ replicas and no additional rounds.
Aequitas~\cite{kelkar2020order} introduced receive-order fairness and identified Condorcet cycles as a fundamental obstacle. Themis~\cite{kelkar2023themis} formalised $\gamma$-batch-order fairness. Compared with Themis, \pname{} uses TEE-attested local orders to prevent equivocation, requiring only
$n > \frac{2\gamma f}{2\gamma-1}$ replicas, and uses synchrony instead of additional replicas to compensate for weaker timing assumptions. It also supports ordering linearizability within the same framework. Other approaches optimize orthogonal aspects while retaining $3f{+}1$ replicas and no trusted hardware, including quick order fairness~\cite{cachin2022quick}, weighted aggregation~\cite{chen2024auncel}, fairness for conflicting transactions~\cite{nagda2024rashnu}, and compact stateless verification~\cite{ren2025autig}. FairDAG-AB/RL~\cite{kang2026fairdagconsensusfairnessmultiproposer}, Tilikum~\cite{segalini2026tilikum}, and Herring~\cite{putnik2026herring} integrate fair ordering into $3f{+}1$ DAG pipelines.

\textbf{Fair ordering with trusted hardware.}
Fairy~\cite{stathakopoulou2021adding} uses $3f{+}1$ TEE-equipped nodes to timestamp encrypted transactions, providing input causality but not ordering linearizability. Aion~\cite{zarbafian2023aion} uses $3f{+}1$ TEEs and challenge--response latency estimation to detect delayed transactions, relying on trusted clocks and stable network delays.
In contrast, \pname{} requires neither trusted clocks nor latency estimation, and uses the known synchronous bound $\Delta$.

\textbf{TEE-assisted BFT.}
A2M~\cite{chun2007a2m} and TrInc~\cite{levin2009trinc} showed that trusted append-only state can prevent equivocation, enabling $2f{+}1$ replication~\cite{veronese2013minbft,kapitza2012cheapbft,behl2017hybster}. Subsequent work addresses responsiveness~\cite{decouchant2022damysus}, trusted-component usage across view changes~\cite{rahli2024oneshot}, rollback attacks~\cite{rahli2026pallas}, and asynchronous DAG-BFT~\cite{xie2025fides}. FlexiTrust~\cite{gupta2023flexitrust} instead retains $3f{+}1$ replicas to avoid responsiveness and rollback limitations.
Trusted-time mechanisms for enclaves~\cite{anwar2019securing,fernandez2023triad,nasrullah2024trusted,hamidy2023t3e,nasrullah2025sok} are orthogonal. \pname{} uses no trusted clock and only requires enclaves to attest monotonically increasing indicators.

\section{Conclusion}
\label{sec:conclusion}

We proved that neither ordering linearizability nor $\gamma$-batch-order fairness with minimal batching is achievable with $n = 2f{+}1$ replicas in partially synchronous networks, even with TEEs. 
However, assuming synchrony removes this impossibility because all correct indicators are then collected within bounded time.
We designed mechanisms to ensure either ordering linearizability with $n = 2f{+}1$ replicas or $\gamma$-batch-order fairness with $n > \tfrac{2\gamma f}{2\gamma-1}$ replicas, below the size of prior systems.
\rev{Our protocol, \pname{}, sustains $3$--$13$ times the throughput of the strongest fair-ordering baseline with a latency of $230$--$300$~ms.}

\section*{Acknowledgment}

ChatGPT and Claude were used as AI/LLM-based assistants for text consolidation and refinement, as well as for coding assistance during the preparation of this work. All AI/LLM-generated or suggested content was manually reviewed and checked by the authors, who remain fully responsible for the accuracy, content and final form of the work. 

\bibliographystyle{IEEEtran}

\end{document}